\documentclass[pra,aps,notitlepage,nofootinbib]{revtex4-1}

\usepackage{bm}
\usepackage{bbm}
\usepackage{epsfig}
\usepackage{wrapfig}
\usepackage{float}
\usepackage{setspace}
\usepackage{mathrsfs}
\usepackage{amsmath,amssymb,amsfonts}
\usepackage{graphicx,xcolor}
\usepackage[a4paper,margin=2.4cm]{geometry}

\usepackage[hypertex, bookmarks=false, pdfstartview=FitH]{hyperref}

\newcommand{\continuousp}[2]{\ensuremath{p_{#1}(#2)}}
\newcommand{\discretep}[2]{\ensuremath{P_{\{#1\}}(#2)}}

\begin{document}

\title{The grasshopper problem}
\author{Olga \surname{Goulko}}
\affiliation{Department of Physics, University of Massachusetts,
  Amherst, MA 01003, USA}
\email{goulko@physics.umass.edu}
\author{Adrian \surname{Kent}}
\affiliation{Centre for Quantum Information and Foundations, DAMTP, Centre for
  Mathematical Sciences, University of Cambridge, Wilberforce Road,
  Cambridge, CB3 0WA, U.K.}
\affiliation{Perimeter Institute for Theoretical Physics, 31 Caroline Street North, Waterloo, ON N2L 2Y5, Canada.}
\email{A.P.A.Kent@damtp.cam.ac.uk} 

\date{\today} 

\begin{abstract}
We introduce and physically motivate the following problem in geometric
combinatorics, originally inspired by analysing Bell inequalities.   
A grasshopper lands at a random point on a planar lawn of area one. It then jumps once, a fixed distance
$d$, in a random direction. What shape should the lawn be to
maximise the chance that the grasshopper remains on the lawn after jumping?  
We show that, perhaps surprisingly, a disc shaped lawn is not optimal
for any $d>0$.
We investigate further by introducing a spin 
model whose ground state corresponds to the solution of 
a discrete version of the grasshopper problem.  
Simulated annealing and parallel tempering searches are consistent
with the hypothesis that
for $ d < \pi^{-1/2}$ the optimal lawn resembles  
a cogwheel with $n$ cogs, where the integer $n$
is close to $ \pi ( \arcsin (  \sqrt{\pi} d /2 ) )^{-1}$. 
We find transitions to other shapes for $d \gtrsim  \pi^{-1/2}$.

\end{abstract}
\maketitle
  
\section{Introduction}

\subsection{Relation to Bell inequalities}

As Bell's celebrated theorem \cite{bell1989einstein} showed, quantum theory is not locally
causal.    The correlations predicted by quantum theory for 
the results of spacelike separated measurements on two or 
more entangled systems cannot be reproduced by any locally
causal model.  Bell inequalities are constraints on correlations, 
for some set of measurements, 
that must be satisfied by any locally causal model but are 
violated when those measurements are carried out on some quantum
states.
Simple Bell inequalities, such as the Clauser-Horne-Shimony-Holt
inequality \cite{clauser1969proposed}, demonstrate the gap between the predictions
of quantum theory and locally causal theories in a form that can be 
experimentally verified, even after allowing for detector inefficiencies and
errors.   

However, the full class of Bell inequalities remains poorly
understood,
even for the simplest case of spin measurements on two spin $1/2$
particles. 
Understanding more general Bell inequalities is intrinsically
interesting, shedding light as it does on the structure and properties
of quantum theory.   There is also some practical motivation, since 
many quantum cryptographic protocols and tests depend on Bell
inequalities to verify that a shared quantum state yields 
quantum correlations of the right type.   This is necessary in
many scenarios since such a state may have been supplied by an 
untrustworthy party or interfered with by an eavesdropper. 
Optimizing protocols requires considering the full class of 
Bell inequalities.  One might expect that inequalities that 
involve completely random measurement choices are good options,
since they give adversaries minimal information.   

With these motivations, Ref.~\cite{kent2014bloch} introduced and analysed Bell
inequalities for the case where two parties carry out spin
measurements about randomly chosen axes and obtain the 
spin correlations for pairs of axes separated by angle 
$\theta$, for each $\theta$ in the range $0 \leq \theta \leq \pi/2$. 
It was shown that the correlations of locally causal models satisfy
bounds violated by quantum theory for $\theta$ in the
range $0 < \theta <  \pi /3 $.
Although the bounds obtained were shown to be optimal for an infinite set of
values of $\theta$ converging to zero, they were not shown to be
optimal for general $\theta$, and indeed we expect that they are not. 
It was noted \cite{kent2014bloch} that obtaining tight bounds would be 
equivalent to solving a geometric combinatorics problem on 
the Bloch sphere.    In its simplest form (which assumes an additional ansatz
about the solution) this problem can be pictured as follows.
Half the area of a sphere is covered by a lawn, with the property that
exactly one of every pair of antipodal points belongs to the lawn.  
A grasshopper lands at a random point on the lawn, and then jumps 
in a random direction through spherical angle $\theta$.   What lawn shape maximises the 
probability that the grasshopper remains on the lawn after jumping,
and what is this maximum probability (as a function of $\theta$)?  

The problem seems harder to solve than to pose.   
Intuitions and sketchy ideas for proofs come readily to mind, but these  
may well be incorrect.\footnote{The lion and man
problem, which asks whether a lion can always catch a man in a bounded
region if they have equal maximum speeds and the man wishes to avoid
capture, gives a well known cautionary example of the pitfalls of intuitive
reasoning: see pp. 114-117 of Ref. \cite{littlewood1986littlewood}.}
One common initial intuition is that the problem has an isoperimetric
flavour.   It appears at first glance that minimizing the length of
the lawn boundary might also minimize the probability of jumping 
outside the lawn, at least for small $\theta$.    If so, a 
hemispherical lawn would be optimal for small $\theta$.   
However, this proof idea neglects the possibility 
that a jump could cross the boundary twice or more.

\subsection{The planar grasshopper problem}

Another intuition is that the problem may be simpler to solve
if translated to the plane.  This means dropping the antipodal
condition, which has no simple analogue for the planar version of the
problem.  Investigating the planar problem thus also tests the 
relevance of the antipodal condition.
A related intuition is that the disc 
should be provably optimal for small jumps in the plane, whether or not there
is an isoperimetric argument for this.   

However, as far as we have been able to establish, not only is
the answer to the planar grasshopper problem not known, but the question does not 
seem to have ever been studied. This seems surprising, given that the question could have been 
posed by Euclid and might plausibly have interested him and
contemporaries.\footnote{There is perhaps some question as to 
how precisely Euclid understood the notion of randomness, although
he certainly used the term; vide Proposition II.4 of Ref.~\cite{euclid}. 
An interesting discussion can be found at 
https://rjlipton.wordpress.com/2016/01/12/did-euclid-really-mean-random/
.}
Remarkably, no even vaguely similar question appears in Croft et al.'s survey
\cite{croftetal}
of unsolved problems in geometry.  The closest results of which we
are aware are those of Bukh \cite{bukh2008measurable}, which 
describe the properties of sets in ${\mathbb R}^n$ that have 
no pair of points separated by any of some finite list of distances.
 
From the perspective of geometric combinatorics, the planar version of
the grasshopper problem seems the most natural starting point.
It is intriguing in its own right, and as we will see, has an
unexpectedly rich structure.   It also tests out intuitions that could 
be formed about the spherical problem, without the extra conditions 
relevant to Bell inequalities.  For example, if a pure 
isoperimetric argument could somehow be run in the spherical case,
without appealing to the extra conditions, intuition suggests a similar
argument should work in the planar case. In particular, if it were possible to show by a purely isoperimetric 
argument that a hemisphere is optimal
on a sphere (of area $2$) for small jump angles $\theta$, then one would also
expect the disc to be optimal in the plane for small jump distances $d$.
However, as we will see, the disc is not optimal in the plane for any
$d>0$. So, any proof idea
for the spherical problem needs either to identify
non-hemispherical optimal solutions and explain why they are
optimal, or to go beyond isoperimetric
intuitions, or to explain why the spherical case is different, or
to explain why the extra conditions matter.  

The planar problem is also the simplest and most natural setting for
exploring experimental approaches to the problem based on
discretizing it and reformulating the discretization in 
terms of statistical
physical spin models, which we describe below. These spin
models are interesting in their own right, as they represent an
unusual class of systems with fixed-range interactions, where the
range can be large. Discretizing the problem also raises
theoretical and computational issues, which are best explored in
the simplest setting.
In this paper, we thus focus on the planar problem.
We explain the numerical techniques used, and the various 
consistency checks that allow reasonable confidence that our 
results are at least qualitatively broadly correct.
We then describe and analyse these results.  
We leave the application of the techniques developed here to the
spherical and other cases for future work.

\section{Statement of the main problem} 

\qquad{\bf Informal statement:} You are given a bag of grass seed from
which you can grow a lawn of any shape (not necessarily connected) with unit area on a
planar surface.  A grasshopper lands at a random point on your lawn,
then jumps a given distance $d$ in a random direction. What lawn
shape should you choose to maximise the probability that
the grasshopper remains on your lawn after jumping? 

The lawn may also be allowed to be of variable density, between zero
and one, with the condition that the integrated density over the plane
is one.  In this case, we take the probability density for the
grasshopper to land at a specific point on the lawn to be the lawn
density at this point.  Likewise we take the probability that the
grasshopper remains on the lawn after jumping to be the lawn density
at its landing point.

\qquad{\bf Formal statement:}  
Consider a probability density
$\mu$ on the plane ${\mathbb R}^2$ satisfying $0 \leq \mu (\mathbf{r})
\leq 1$ for all $\mathbf{r} \in {\mathbb R}^2$ and
\begin{equation}
\int_{{\mathbb R}^2}d^2r \mu  (\mathbf{r}) = 1 \, . 
\label{eq:munorm}
\end{equation}
 The functional $\continuousp{\mu}{d}$ is 
defined by 
\begin{eqnarray}\label{successprob}
  \continuousp{\mu}{d} &=& \frac{1}{ 2 \pi d} 
\int_{{\mathbb R}^2}d^2r_1 \int_{{\mathbb R}^2}d^2r_2 \mu(\mathbf{r}_1) \mu( \mathbf{r}_2 )  \delta(|\mathbf{r}_1-\mathbf{r}_2|-d)\\
  &=& \frac{1}{ 2 \pi} \int_{{\mathbb R}^2}d^2r \int_{0}^{2\pi}d \theta
  \mu(\mathbf{r}) \mu( \mathbf{r} - d \, \hat{n}(\theta) ) \,
  , \nonumber
\end{eqnarray} 
with the unit vector $\hat{n}(\theta) = ( \cos \theta , \sin \theta
)$. What is the supremum of $\continuousp{\mu}{d}$ over all such
density functions $\mu$, for each value of $d>0$?  Which
$\mu$, if any, attain the supremum?  Or if none, which
sequences approach the supremum value?

For most of this paper we focus on the case where the lawn density $\mu$
takes only the values $0$ or $1$, i.e.\ the lawn is a shape of area $1$
and uniform density.  This is a natural version of the problem.
Moreover, as we explain below, investigating it also gives a great
deal of insight into the general case.

\qquad{\bf Formulation in Fourier space:} It is illuminating to
recast the functional $\continuousp{\mu}{d}$ defined in
  Eq.~(\ref{successprob}) in terms of the Fourier transform of the
  density function $\mu$, defined via $\mu(\mathbf{r}) =
  \int_{\mathbb{R}^2}d^2p\tilde{\mu}(\mathbf{p})e^{i\mathbf{p}\mathbf{r}}/(2\pi)^2$
  and $\tilde{\mu}(\mathbf{p})=\int_{\mathbb{R}^2}d^2r
  \mu(\mathbf{r})e^{-i\mathbf{p}\mathbf{r}}$. The normalisation
  condition~\eqref{eq:munorm} then implies
  $\tilde{\mu}(p=0)=1$. In this representation the functional is
\begin{equation}
\continuousp{\tilde{\mu}}{d}=\frac{1}{(2\pi)^2}\int_{\mathbb{R}^2}d^2p |\tilde{\mu}(\mathbf{p})|^2J_0(dp),
\label{fourierspaceprob}
\end{equation} 
where $J_\alpha(z)$ is the Bessel function of the first kind. Note
that if $\mu(\mathbf{r})$ is an even function, then
$\tilde{\mu}(\mathbf{p})$ is real and even and if $\mu(\mathbf{r})$ is
radially symmetric then $\tilde{\mu}(\mathbf{p})$ is real and radially
symmetric also.

\qquad{\bf Success probability for the unit disc lawn:} The success
probability for lawns in the shape of the unit disc (with density one
inside the disc and zero outside) can be calculated exactly by
analytically solving the integral (\ref{successprob}) for $\continuousp{\mu}{d}$,
or equivalently the integral (\ref{fourierspaceprob}) for $\continuousp{\tilde{\mu}}{d}$,
\begin{equation}
\continuousp{\rm disc}{d}=1-\frac{2}{\pi}\left(\frac{d\sqrt{\pi}}{2}\sqrt{1-\frac{d^2\pi}{4}}+\arcsin\left(\frac{d\sqrt{\pi}}{2}\right)\right),
\label{eq:unitdiscprob}
\end{equation}
assuming $d\leq\pi^{-1/2}$. 
Its Taylor expansion for small $d$ is
\begin{equation}
\continuousp{\rm disc}{d}=1- \frac{2}{\sqrt{\pi}} d + \frac{\sqrt{\pi}}{12} d^3 + O (d^5 ) \,. 
\end{equation} 
The success probability approaches one as $d\rightarrow0$ and decreases monotonically with increasing $d$.

\section{Analytic results} 
\label{sec:analytical}
{\bf Lemma 1:}  \qquad The disc of area $1$ (radius $\pi^{-1/2}$) is not optimal for any $d >
\pi^{-1/2}$.  

{\bf Proof:} \qquad  Suppose $d > \pi^{-1/2}$ and suppose that the
lawn is the disc $D$ of area $1$ centred at the origin.  If the grasshopper lands within the 
disc $D_d$ of radius $r_d = \min ( \pi^{-1/2}, d -   \pi^{-1/2} )$ centred at the 
origin, then its jump will take it outside $D$ with probability $1$. 
Removing $D_d$ from $D$ and redistributing its area to some shape $S$ outside $D$ thus
cannot lower the overall success probability.   Moreover, it is easy
to find shapes $S$ that increase the overall success probability.  
For (a far from optimal) example, one could take $S$ to be a rectangle, lying
anywhere outside $D$, 
of width $\epsilon$ and height $h=\pi r_d^2 \epsilon^{-1}$, where $\epsilon$ is chosen so that $h > d$. 

{\bf Comment:} \qquad  This argument clearly fails for $d <
\pi^{-1/2}$, suggesting that we might expect different types of
solution in the regimes $d< \pi^{-1/2}$ and $d > \pi^{-1/2}$.  
The construction also suggests that solutions for $d > \pi^{-1/2}$
might fail to be simply connected, since it begins by creating a hole
in the centre of the disc.  In fact, we show below that 
the disc is not optimal even for $d< \pi^{-1/2}$, so that the 
construction does not directly apply and does not necessarily imply 
a transition at precisely $d = \pi^{-1/2}$.  However, our results also
suggest that for $d$ above and close to $\pi^{-1/2}$ the optimal solution is not simply connected, or
even connected.

{\bf Lemma 2:} \qquad There exists an infinite sequence of jumps
$(d_n)$, with $0<d_{n+1}<d_n$ for all positive integer
$n$ and $\lim_{n\rightarrow\infty}d_n \rightarrow 0$, such
that for each $d_n$ the disc of area $1$ is not optimal.

{\bf Comment:} \qquad The intuition that the disc should
be optimal for sufficiently small $d$ is thus incorrect.  

{\bf Proof:} \qquad Let $d_n = 2 \pi^{-1/2} \sin ( \pi / n )$ for $n\geq2$.
This gives $d_n > \pi^{-1/2}$ for $2\leq n<6$, so Lemma $1$ already shows
the disc is not optimal for these cases. Note that for $n\geq3$, $d_n$
is the edge length of a polygon with $n$ edges inscribed in the unit
disc. 
For $n \geq 6$, choose parameters $a,b$ such that 
$a>b>0$ and $a+b = 2 \pi^{1/2} n^{-1}$, so that the 
perimeter of the unit disc is $n ( a+ b)$.
Take $\epsilon >0$ to be small.   

\begin{figure}
\includegraphics[width=0.4\textwidth]{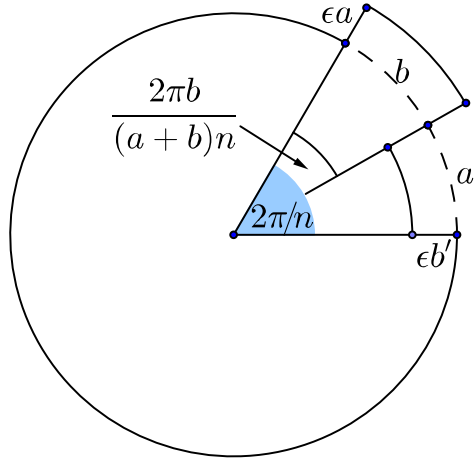}
\caption{\label{fig:proofsketch}Construction of the shape $D_{a,b,n,\epsilon}$ from the disc.}
\end{figure}
A shape $D_{a,b,n,\epsilon}$ of area $1$ can be constructed as
follows, see Fig.~\ref{fig:proofsketch}.   In polar coordinates $(r , \theta)$, $D_{a,b,n,\epsilon}$
is defined by the disc segments 
\begin{eqnarray}
0 \leq r \leq \pi^{-1/2} + \epsilon a &\ \ {\rm for}\ \ & 2\pi k/n  \leq \theta
\leq 2\pi k/n + \frac{2\pi b}{(a+b)n} \, \\
0 \leq r \leq \pi^{-1/2}  - \epsilon b' & {\rm for} & 2\pi k/n + \frac{2\pi b}{(a+b)n}  \leq \theta
\leq 2\pi(k+1)/n  \, , \nonumber 
\end{eqnarray} 
for integer $k$ in the range $0 \leq k \leq n-1$. Here $b'$ is determined by the area $1$ condition, which gives us 
\begin{equation}
b' = \epsilon b + \epsilon^2 \frac{\pi^{1/2} }{2} (a+b )b + {\rm O}(
\epsilon^3 ) \, .
\end{equation}
Fix $n$ and then fix $a,b$ satisfying the conditions above. Now, since $\epsilon$ is small, we can consider $D_{a,b,n,\epsilon}$
as a close approximation to $D$, and calculate the difference 
in their success probabilities to lowest non-zero order in $\epsilon$;
it turns out we do not need the higher order terms in $b'$.  

Calculating the difference in the relevant integrals gives that 
the difference in success probabilities is
\begin{equation}
p( D_{a,b,n,\epsilon} ) - p(D) = n d^2 \left( 1 - \frac{ \pi d^2 }{4}
\right)^{-1/2} \pi^{1/2} \epsilon^2 a b ( a - b) + 
{\rm O}(
\epsilon^3 ) \, .
\label{eq:pdabneps}
\end{equation}  
For $a>b$ and sufficiently small $\epsilon$ this is positive.  Hence
the disc $D$ is also not optimal for $d=d_n$ when $n \geq 6$.

{\bf Comment:} \qquad We see that perturbing the disc by alternating
thin protrusions and thicker indentations, producing a shape with
$n$-fold rotational symmetry, gives a higher success
probability when 
\begin{equation}
d \approx 2 \pi^{-1/2} \sin ( \pi / n )  \,
, \label{eq:nfoldsymm}
\end{equation} 
and so $d \approx 2 \pi^{1/2} n^{-1}$ for large $n$.
In fact, our numerical results from statistical modelling are consistent with the
hypothesis that the optimal shape has $n$-fold symmetry, approximately
related to $d$ as above, for all $d < \pi^{- 1/2}$.    We  
describe these results in Sec.~\ref{sec:cogwheelregime}.

{\bf Theorem 3:} \qquad The disc of area $1$ is not optimal 
for any $d>0$.   

{\bf Proof:} \qquad For $d> \pi^{-1/2}$ this is already established by
Lemma $1$ and for $d = \pi^{-1/2}$ it is established by Lemma $2$.  To
prove that the disc is not optimal for $d$ in the range $0 < d <
\pi^{-1/2}$ we consider $d_{m,n} = 2 \pi^{-1/2} \sin ( m \pi / n )$
for positive integers $m,n$ with no common divisor such that $m \geq
2$ and $ \frac{m}{n} < \frac{1}{6}$.  In this construction, $d_{m,n}$
is the edge length of an $\{n/m\}$ regular stellated polygon, where $n$ is the
number of vertices and $m$ is the winding number around the
centre. Hence there are $(m-1)$ vertices between each two
vertices separated by the edge of length $d$. 

As before, choose parameters $a,b$ such that 
$a>b>0$ and $a+b = 2 \pi^{1/2} n^{-1}$, so that the 
perimeter of the unit disc is $n ( a+ b)$, and take $\epsilon >0$ to be small. A shape $D_{a,b,n,\epsilon}$ of area $1$ can be constructed as
in the proof of Lemma 2.
Fix $n$ and then fix $a,b$ satisfying the conditions above. 
As in the proof of Lemma 2, we take $\epsilon$ small, so that $D_{a,b,n,\epsilon}$
is a close approximation to $D$.  The difference 
in their success probabilities to lowest non-zero order in
$\epsilon$ is also given by expression \eqref{eq:pdabneps},
as before. 

For $a>b$ and sufficiently small $\epsilon$ the probability difference
\eqref{eq:pdabneps} is positive and hence
the disc $D$ is not optimal for $d=d_{m,n}$.  
By continuity, there is a neighbourhood of each $d=d_{m,n}$ for which
the disc is also not optimal.   But the $d_{m,n}$ form a dense subset
of the interval $ \lbrack 0 , \pi^{-1/2} \rbrack $.  
Hence the disc is not optimal for any $d>0$. 

{\bf Comment:} \qquad Lawn shapes with $n$ protrusions and
indentations can be constructed as described above, where $n,m$ 
approximately solve
\begin{equation}
d = 2 \pi^{-1/2} \sin ( m \pi / n ) \, . 
\label{eq:nmfoldsymm}
\end{equation}
We can thus approximate a given value of $d$ either by
\eqref{eq:nfoldsymm} or (generally more closely) by
\eqref{eq:nmfoldsymm} with $m \geq 2$.  Our argument implies that, for
that value of $d$ and some approximation \eqref{eq:nmfoldsymm}, there
is a shape with the symmetries of the $\{n/m\}$ stellated polygon that
has higher probability than the disc.  However, it does not imply that
any such shape is necessarily optimal.  Numerically the best shapes
found have a $n$-fold degree of symmetry, where the integer $n$ is
close to the solution of \eqref{eq:nfoldsymm} and hence smaller than
the $n$ obtained from the stellated construction
\eqref{eq:nmfoldsymm}. This was seen numerically even in the cases
where the solution of \eqref{eq:nfoldsymm} was close to a
half-integer. In such cases both the next-highest and the next-lowest
integer approximations give a higher success probability than the
solution from \eqref{eq:nmfoldsymm}.  We describe these results in
Sec.~\ref{sec:cogwheelregime}; see in particular
Fig.~\ref{fig:optcogsandenergies}.

{\bf Comment:} \qquad Suppose that the optimal solution for jump $d$ is given by 
a density one lawn whose shape $S$ has a smooth boundary.   Consider
two points $x_1 , x_2$ on the boundary of $S$ and the circles
$C_i$ of radius $d$ centred at $x_i$. Let $S_i = S \cap C_i$, and let $l_i$ be the 
sum of the lengths of the arcs comprising $S_i$, for $i=1,2$.  

We now give an unrigorous argument that $l_1 = l_2$. Suppose this is not the case,
and without loss of generality that $l_1 > l_2$.   Then, by
continuity, analogous inequalities hold for points (inside or outside $S$) 
in small neighbourhoods of $x_1$ and $x_2$. 
Subtracting a small area from $S$ that includes $x_2$ and adding the 
same area to the exterior of the boundary of $S$ around $x_1$ then 
should increase the success probability, which would contradict the
optimality of $S$.   

This observation suggests what we will call the equal arc length
hypothesis \cite{lentnerprivatecomm}: for any given $d$, the sum of the arc lengths for $S_i$ is identical
for all points $x_i$ on the boundary of the optimal shape $S$.
Of course, the disc satisfies the equal arc length hypothesis for all
$d$. The best shapes found numerically also appear approximately consistent with
the hypothesis, although the current numerical resolution is not high
enough to give compelling evidence.  We leave further investigation 
for future work.  

\section{Spin model for the discrete grasshopper problem}

We define a discrete version of the grasshopper problem by
dividing the plane into grid cells and assigning a spin
variable $s_i$ to the centre of each cell $i$, where $s_i = 0$ and
$s_i = 1$ correspond to the cell being unoccupied or occupied by the
lawn, respectively. Here we only consider regular lattices, for which
all cells have the same area $A$. The area one condition for the lawn
then corresponds to a fixed number of spins having value $1$, so that
we have $\sum_{i}s_i=N=1/A$.

If we treat a discretized lawn as simply a restricted case of the
general problem, its success probability is defined by
Eq.~(\ref{successprob}). However, we aim to identify approximate
solutions to the grasshopper problem by using simple statistical
physics models that involve quantities that can be computed more
quickly. To do this, we discretize the functional in
Eq.~(\ref{successprob}), such that in the continuum limit
($A\rightarrow0$ and $N\rightarrow\infty$, keeping the total area
$NA=1$ fixed) the discrete functional approaches the exact continuous
expression. We have the following correspondence between continuous
and lattice quantities:
\begin{eqnarray}
\int_{\mathbb{R}^2}d^2r &\leftrightarrow& A\sum_{\mathbf{r}_i\in L},\\
\mu (\mathbf{r}) &\leftrightarrow& \mu (\mathbf{r}_i)=s_i, \\
\delta(x-x_0) &\leftrightarrow& \delta_h(x-x_0).
\end{eqnarray}
Here $\mathbf{r}_i$ are the positions of the sites of the lattice $L$ and $h$ is
the lattice spacing, which scales like $h\propto\sqrt{A}=1/\sqrt{N}$
(for a square lattice $A=h^2$). The discretization involves a
smoothed approximation $\delta_h$ to the (one-dimensional)
$\delta$-function, which appears in (\ref{successprob}). We can
construct $\delta_h$ through a continuous function $\phi$ with finite
support and unit integral in the following manner,
\begin{equation}
\delta_h(x-x_0)=\frac{1}{h}\phi\left(\frac{x-x_0}{h}\right).
\end{equation}
With this definition $\delta_h(x-x_0)\rightarrow\delta(x-x_0)$ as
$h\rightarrow0$. The function $\phi$ must fulfill certain additional
conditions, which are introduced and motivated in Ref.~\cite{peskin2002deltafn}, to compensate as much as possible
for the inaccuracy associated with the presence of the discrete
grid. Here we use two alternative definitions of $\phi$, given by
Eqs.~(\ref{eq:discretedistancefn}, \ref{eq:discretedistancefn2}),
extensive tests of which can be found in Ref. \cite{yangetal2009deltafn}.

The discrete version of the functional $\continuousp{\mu}{d}$ then becomes
\begin{equation}
\discretep{s}{d} = \frac{1}{2\pi d N^{2}h}\sum_{i, j}  s_i s_j \phi \left( \frac{ | \mathbf{r}_i -\mathbf{r}_j  | - d}{h} \right) \, . \label{eq:spinham}
\end{equation}
In the continuum limit, $\discretep{s}{d}\rightarrow \continuousp{\mu}{d}$. To resolve
  the jump distance $d$ we require $h \ll d$. We look for spin
  configurations that maximise $\discretep{s}{d}$, which represent solutions to the
  discrete grasshopper problem considered.

This problem is equivalent to finding the ground state of a spin system with
Hamiltonian $H=-\discretep{s}{d}$. The range of the interactions for this
Hamiltonian is fixed, but not nearest neighbour. 
It is thus an Ising model  with fixed
total spin \cite{newman1999annealingbook} and 
with fixed-range attractive interactions. The interaction range
is approximately the grasshopper jump distance $d$, modulo small
variations introduced by the discretization of the delta function. 
To the best of our knowledge, systems 
with interactions that have a fixed range significantly larger
than the lattice spacing have not been studied before.
Our results suggest this class of statistical models have some
interesting and unusual properties.
   
\section{Numerical Setup}
Our discretization involves theoretical choices: the type of lattice,
the number of occupied sites, and the precise form of the
$\delta$-function approximation $\phi$. The default setup is a
square grid with fixed total spin $N$ and grid cell size (lattice
spacing) $h$ determined by the area one condition: $h^2N=1$. We varied
the lattice spacing $h$ by performing simulations for several
different $N$ between 5000 and 90000, corresponding to $h$ between approximately 0.015 and 0.003. As the default, we
use the following approximation to the $\delta$-function
\cite{peskin2002deltafn},
\begin{equation}
\phi_1 \left(  \frac{\Delta x}{h} \right) = \left\{
\begin{array}{ll}
\frac{1}{4}\left(1+\cos(\frac{\pi\Delta x}{2h})\right) & {\rm if~} |\Delta x|/h\leq 2 \\
0 & {\rm if~} |\Delta x|/h\geq 2
\end{array}.\right.\label{eq:discretedistancefn}
\end{equation}
This means that two occupied lattice sites contribute to $\discretep{s}{d}$ if the absolute difference between their distance and $d$
is at most $2$ lattice spacings. The contribution is larger the closer
the distance is to $d$. To assess the effect of the $\delta$-function
discretization we also performed calculations with a different choice
for $\phi$, discussed in \cite{yangetal2009deltafn},
\begin{equation} \renewcommand\arraystretch{1.8} \phi_2 \left(
\frac{\Delta x}{h} \right) = \left\{
\begin{array}{lll} \frac{17}{48} & +
\frac{\sqrt{3}\pi}{108}+\frac{|\Delta x|}{4h} - \frac{\Delta
x^2}{4h^2} + \frac{1-2|\Delta x|/h}{16}\sqrt{1 + \frac{12|\Delta
x|}{h}-\frac{12\Delta x^2}{h^2}} & \\ & -
\frac{\sqrt{3}}{12}\arcsin\left(\frac{\sqrt{3}|\Delta
x|}{h}-\frac{\sqrt{3}}{2}\right) & {\rm if~} |\Delta x|/h\leq 1, \\
\frac{55}{48} & -\frac{\sqrt{3}\pi}{108}-\frac{13|\Delta x|}{12h} +
\frac{\Delta x^2}{4h^2} + \frac{2|\Delta x|/h - 3}{48}\sqrt{-
23+\frac{36|\Delta x|}{h} - \frac{12\Delta x^2}{h^2}} & \\ & +
\frac{\sqrt{3}}{36}\arcsin\left(\frac{\sqrt{3}|\Delta
x|}{h}-\frac{3\sqrt{3}}{2}\right) & {\rm if~} 1\leq|\Delta x|/h\leq 2,
\\ 0 & & {\rm if~} |\Delta x|/h\geq 2.
\end{array}\right.\label{eq:discretedistancefn2}
\end{equation} To check whether the structure of the lattice affects
the optimal lawn shape, we also simulated the model on a hexagonal
grid. In this case the area of a grid cell is $A=1/N=\sqrt{3}h^2/2$,
where the lattice spacing $h$ is the distance between the centres of
the hexagonal cells.

\begin{figure}
\includegraphics[clip=true,width=0.49\textwidth]{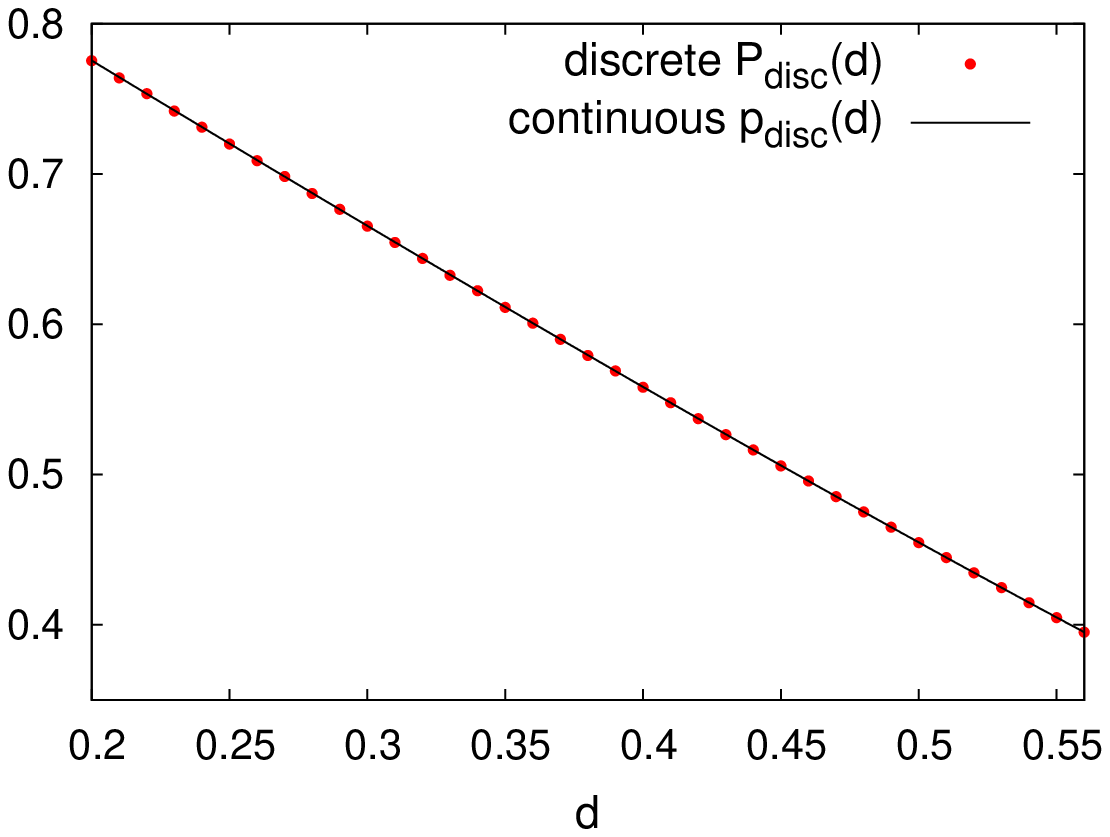}\hfill
\includegraphics[clip=true,width=0.49\textwidth]{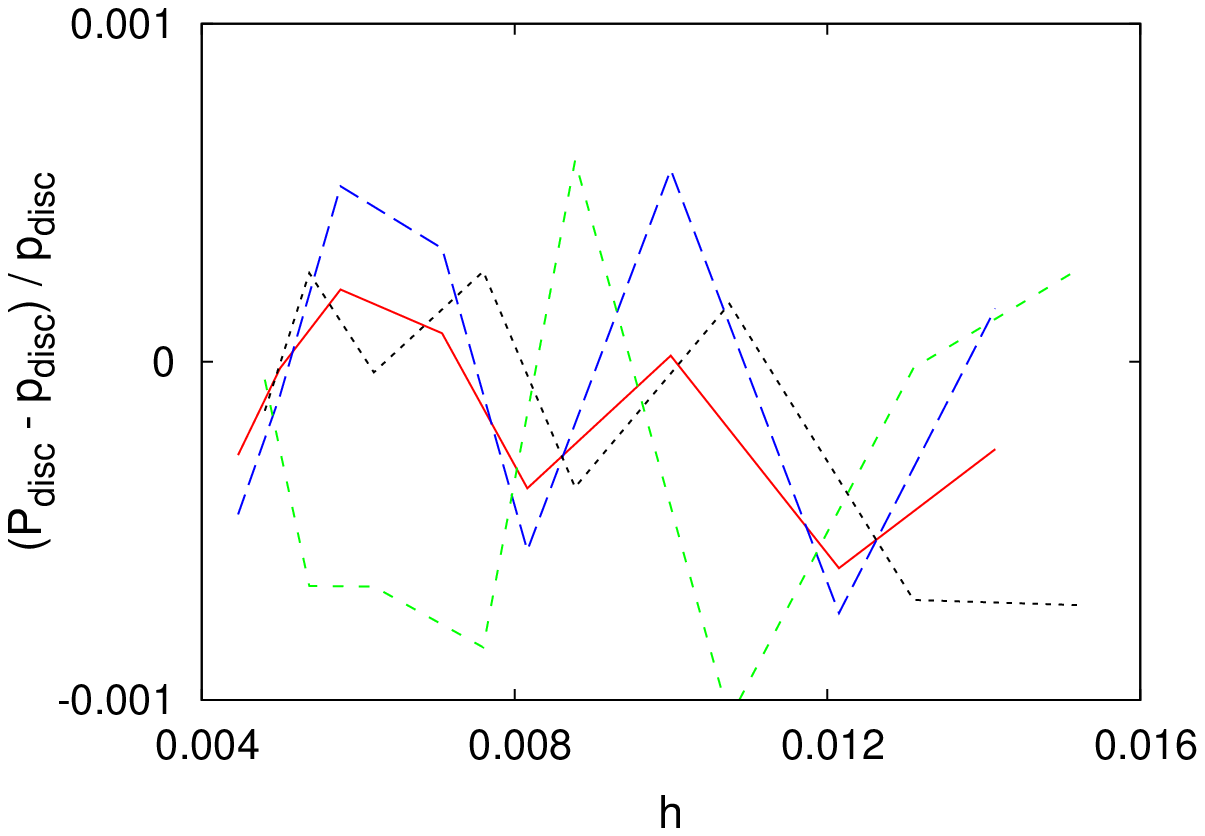}
\caption{\label{fig:consistencychecks}Study of discretization
 effects. Left panel: The exact continuous probability functional
  $\continuousp{\mu}{d}$ for the unit disc given by Eq.~\eqref{eq:unitdiscprob} (solid line) compared with the discrete $\discretep{s}{d}$ for the unit disc (red dots) for $0.2\leq d\leq0.56$. Right panel:
  Relative deviation of $\discretep{s}{d}$ for the unit disc configuration from
  $\continuousp{\mu}{d}$ for the unit disc as function of the lattice spacing
  $h$ for $d=0.27$ and for several discretization setups: square grid with
  $\phi=\phi_1$ (red solid line); square grid with $\phi=\phi_2$
  (blue long dashed line); hexagonal grid with $\phi=\phi_1$ (green
  short dashed line) and hexagonal grid with $\phi=\phi_2$ (black
  dotted line). Lines are to guide the eye.}
\end{figure}

To illustrate the effects of discretization and the specific setup,
Fig.~\ref{fig:consistencychecks} shows the dependence of $\discretep{s}{d}$ for the
unit disc on the discretization parameters, in comparison to the exact
continuous model probability~(\ref{eq:unitdiscprob}). For all setups
considered the results agree well with each other and the deviation
from the continuous solution is of order $0.1\%$ or 
less. Discretization effects for the best configurations found numerically are discussed in Sec.~\ref{sec:cogwheelregime}.

We use simulated annealing \cite{KiGeVe83} and parallel tempering
\cite{Hukushima1996PT} to find configurations that maximise
$\discretep{s}{d}$ for different values of $d$. For
this we introduce an additional simulation parameter $T$, which is the
analogue of a temperature. At fixed temperature, in each simulation
step the spin configuration is updated as follows
\cite{newman1999annealingbook}. We select a random lattice site $i$
with $s_i=0$ and a random lattice site $j$ with $s_j=1$. We then
attempt to exchange the spin values such that $s'_i=1$ and
$s'_j=0$. The exchange is accepted with the Metropolis acceptance
probability $p=\min (1, e^{\Delta P/T})$ and rejected with probability
$(1-p)$. Here $\Delta P=\discretep{s'}{d}-\discretep{s}{d}$ is the difference between the values of the new and the old configuration. If $\Delta
P>0$ the update is
deterministically accepted, otherwise it can still be accepted with a
probability that exponentially decreases as a function of $\Delta
P$. The higher the temperature, the likelier are we to accept an
update that decreases $\discretep{s}{d}$.  At low temperatures, when the
acceptance rate is low, it can be advantageous to use continuous time
Monte Carlo updates instead. In this case the pair of lattice sites is
selected according to their pre-calculated exchange probability,
rather than uniformly, and the exchange is then deterministically
accepted \cite{newman1999annealingbook}. Since here we are only
interested in the final zero temperature state, we do not need to keep
track of the Monte Carlo time.

For a simple simulated annealing search, we start the simulation with
a random spin configuration at high temperature and then gradually
decrease the temperature until a minimal temperature is reached. We
use an exponential cooling schedule: at each annealing round the
temperature is multiplied by a constant factor $0<\alpha<1$. The
number of simulation steps between the annealing rounds must be large
enough to allow the system to reach a stationary state. 
Several animations showing examples of the annealing process can be found in the Supplemental Material.
While simulated annealing is effective at
identifying global extrema for many systems, it can fail,
particularly for rugged energy landscapes.

In parallel tempering, also known as replica exchange Monte
Carlo, several copies of the system are simulated in parallel, each
at a different fixed temperature.   After some number of steps,
exchange updates attempting to swap configurations at neighboring
temperatures are suggested. The acceptance probability for the swap
updates is min$(1,e^{\Delta P\Delta\beta})$, where $\beta=1/T$ is the
inverse temperature. This method is generally more successful in
probing complex energy landscapes, since through exchanging
configurations with those at higher temperatures the system can escape
from local extrema. Nevertheless, no statistical method is 
guaranteed to find a global extremum.

We performed extensive tests to find optimal annealing and parallel
tempering schedules and performed several independent simulations for
each set of parameters. To establish the best shape for values of $d$
for which our results were ambiguous, for instance near a transition
between two different types of shape with similar values of $\discretep{s}{d}$, we performed
additional checks: each potentially optimal solution was set as the
initial configuration, and the annealing process was then run with a
sufficiently low starting temperature to ensure that the rough shape
was preserved, while attempting to further optimize $\discretep{s}{d}$. 
\section{Numerical results}\label{numres}
All results presented in this section were obtained with a square
grid, $N=10000$ and the $\delta$-function approximation $\phi=\phi_1$
given by Eq.~(\ref{eq:discretedistancefn}), unless stated otherwise.  Checks were
also performed with alternative setups (hexagonal grid, $\phi=\phi_2$, different values of $N$ up to $N=90000$) and produced consistent
results.

\subsection{Cogwheel regime}
\label{sec:cogwheelregime}
\begin{figure}
\includegraphics[clip=true,width=0.16\textwidth]{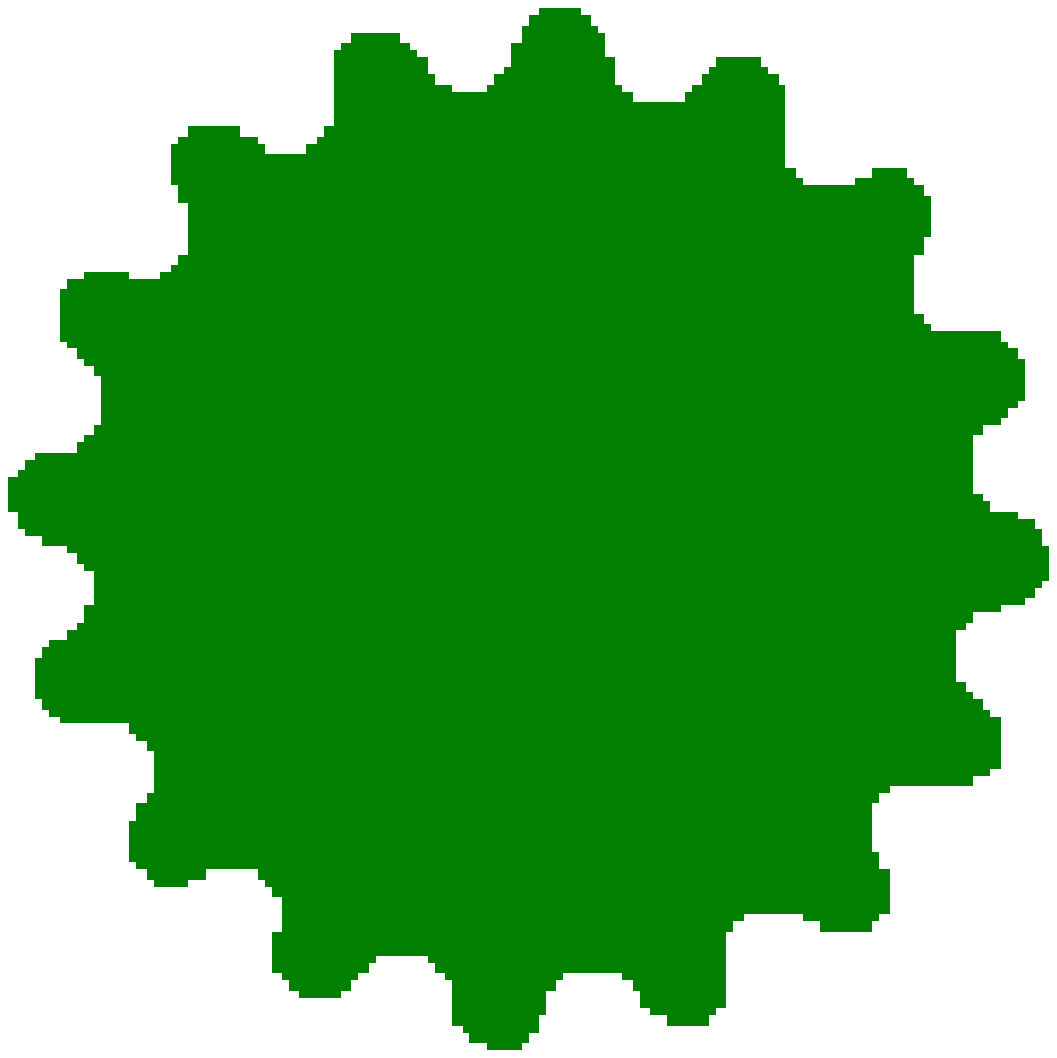}\hfill
\includegraphics[clip=true,width=0.16\textwidth]{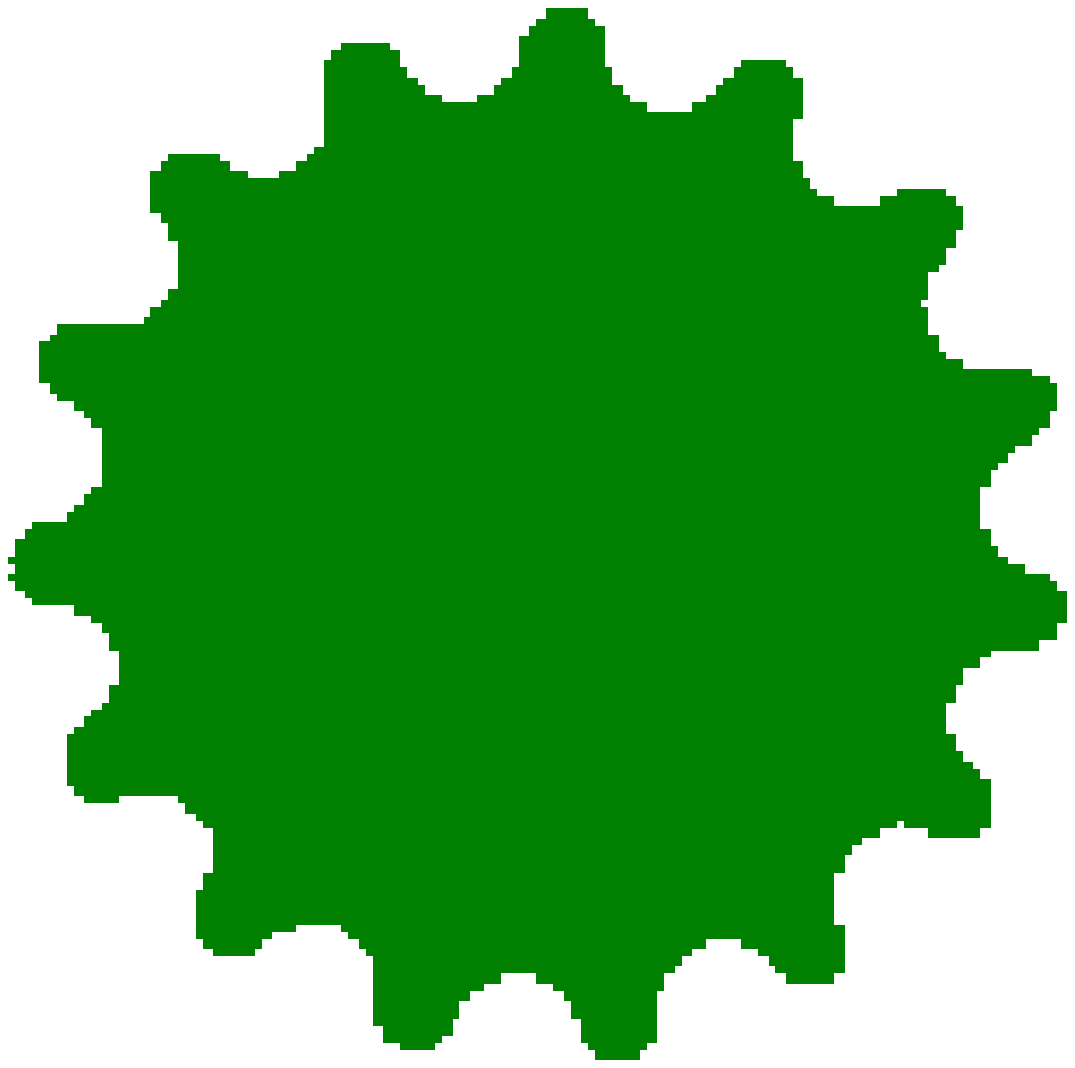}\hfill
\includegraphics[clip=true,width=0.16\textwidth]{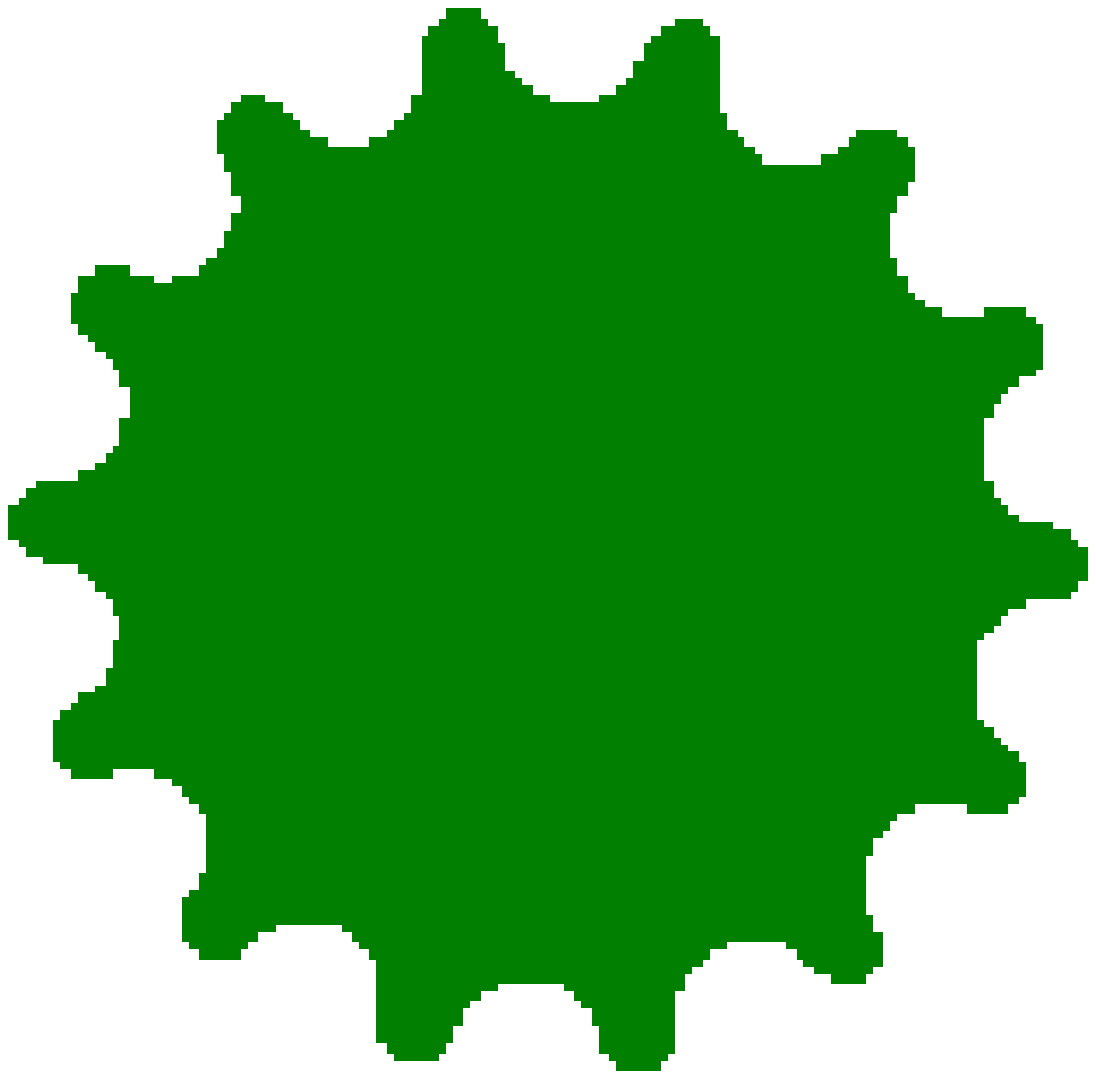}\hfill
\includegraphics[clip=true,width=0.16\textwidth]{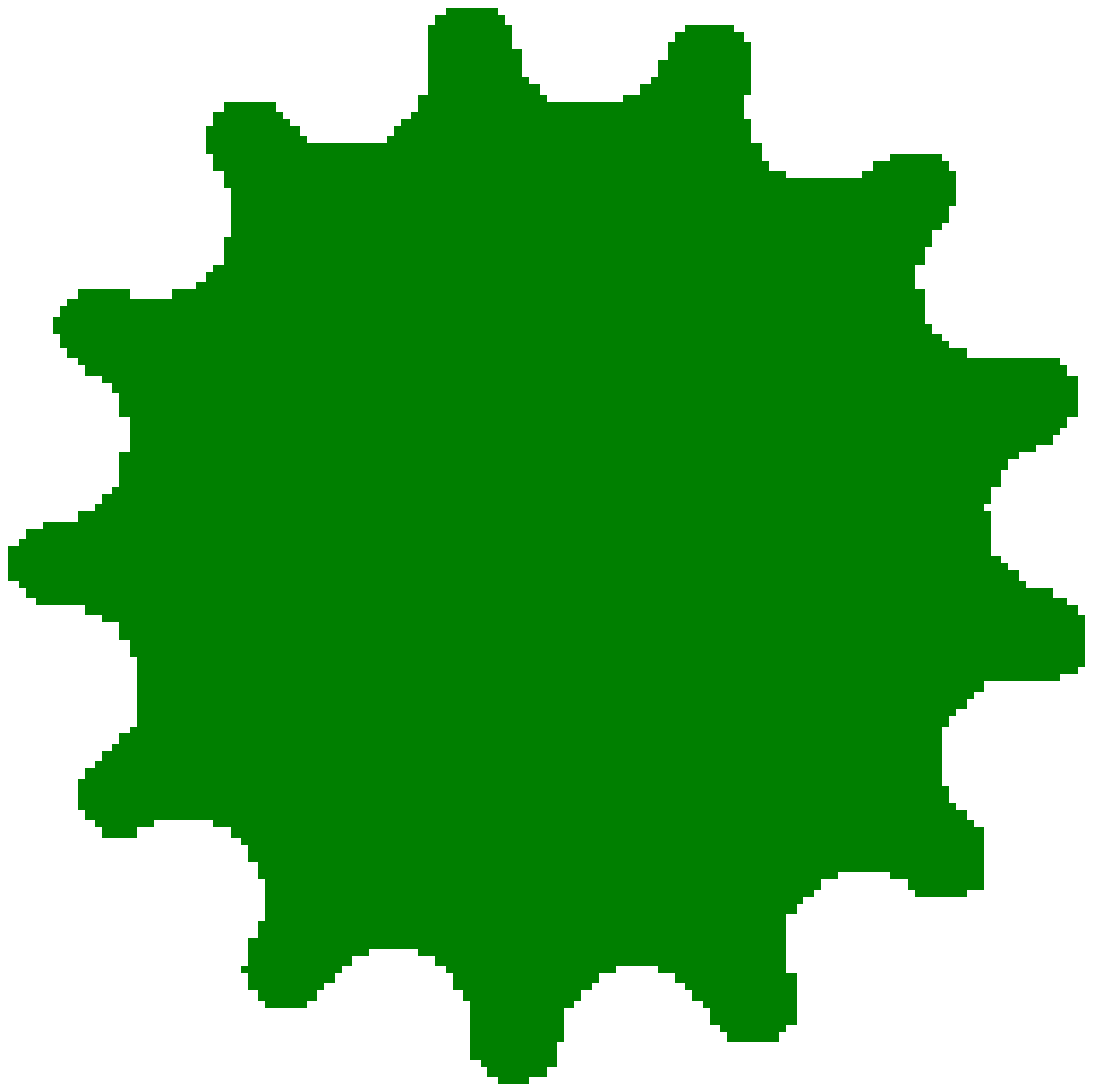}\hfill
\includegraphics[clip=true,width=0.16\textwidth]{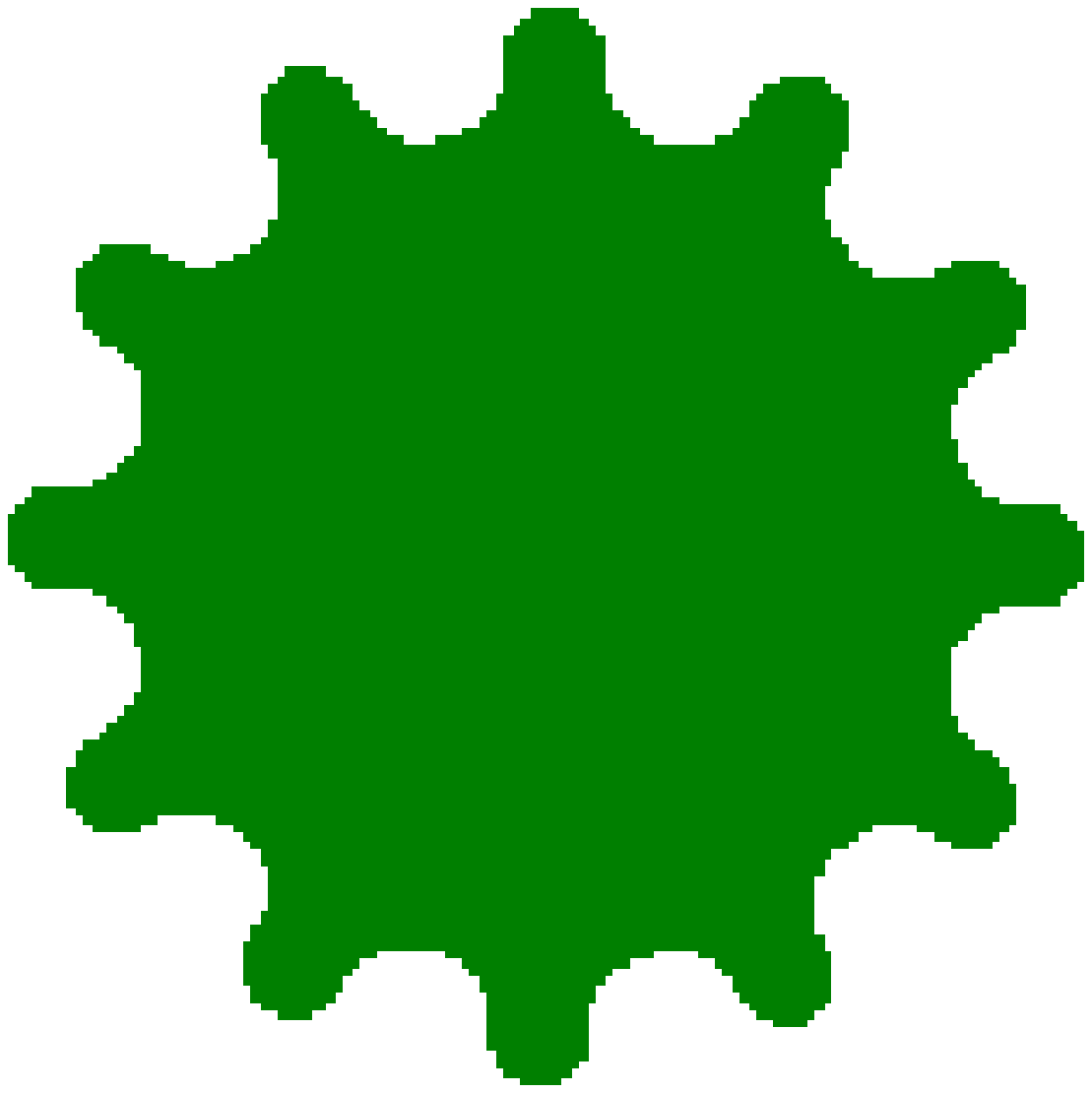}\hfill
\includegraphics[clip=true,width=0.16\textwidth]{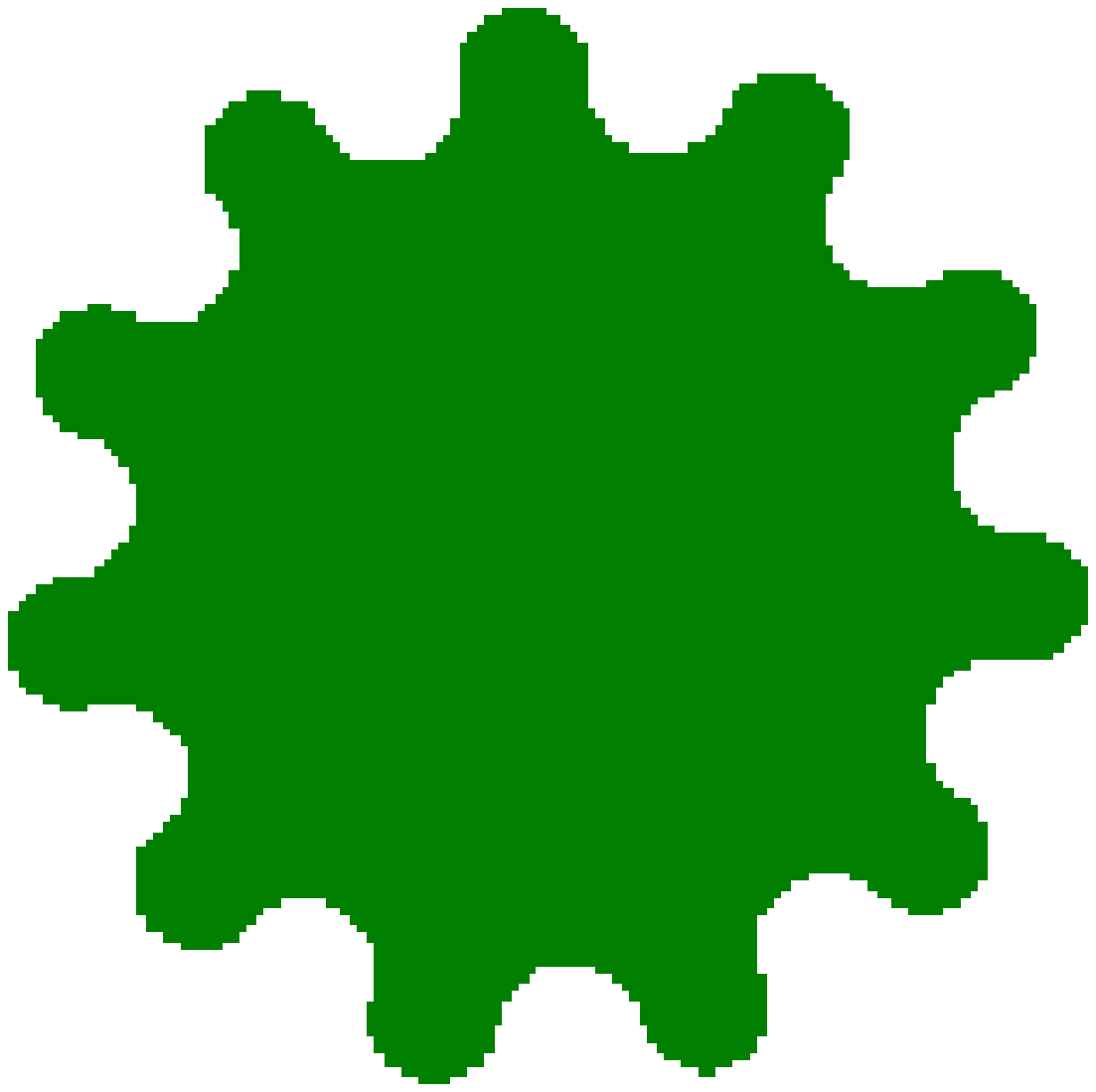}\hfill
\includegraphics[clip=true,width=0.16\textwidth]{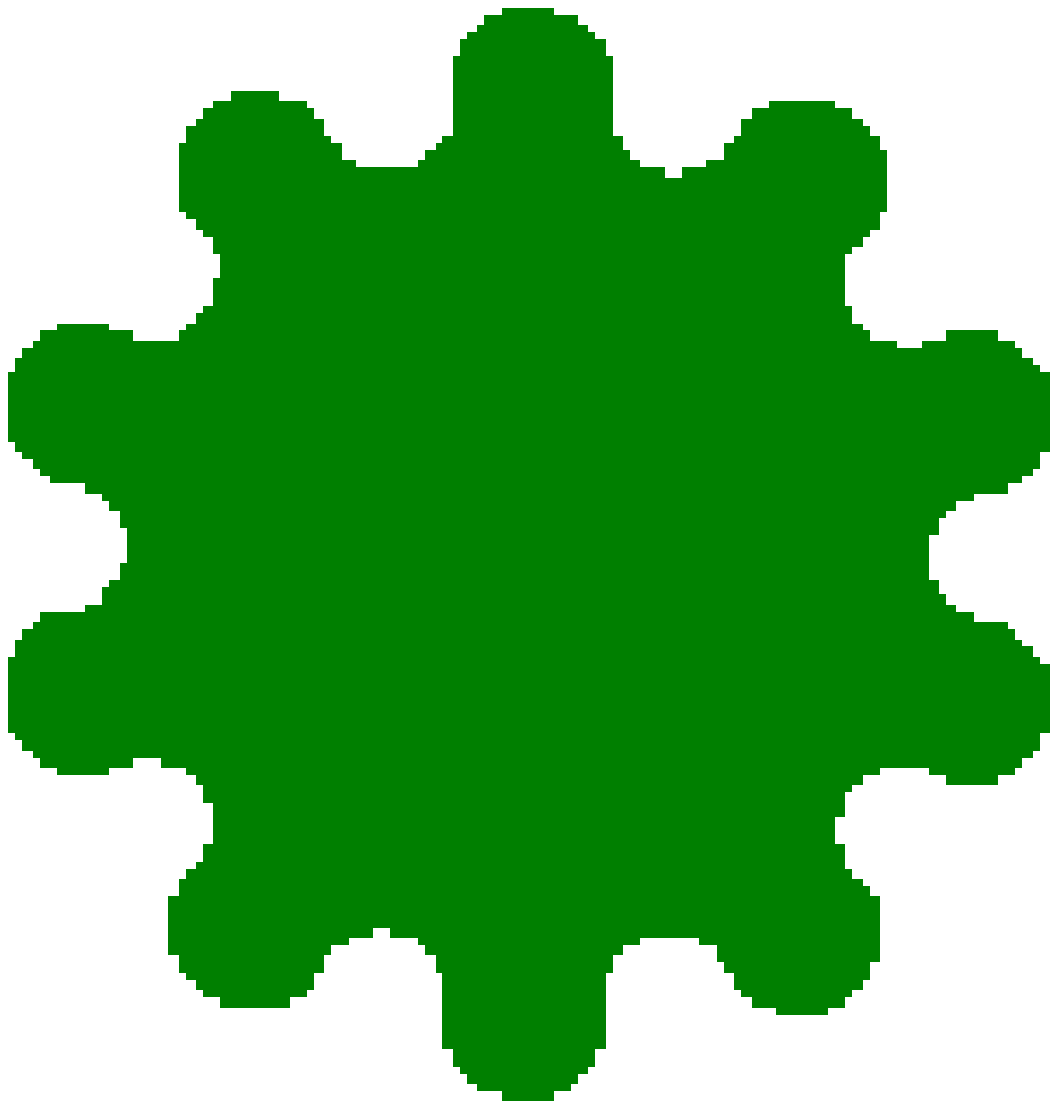}\hfill
\includegraphics[clip=true,width=0.16\textwidth]{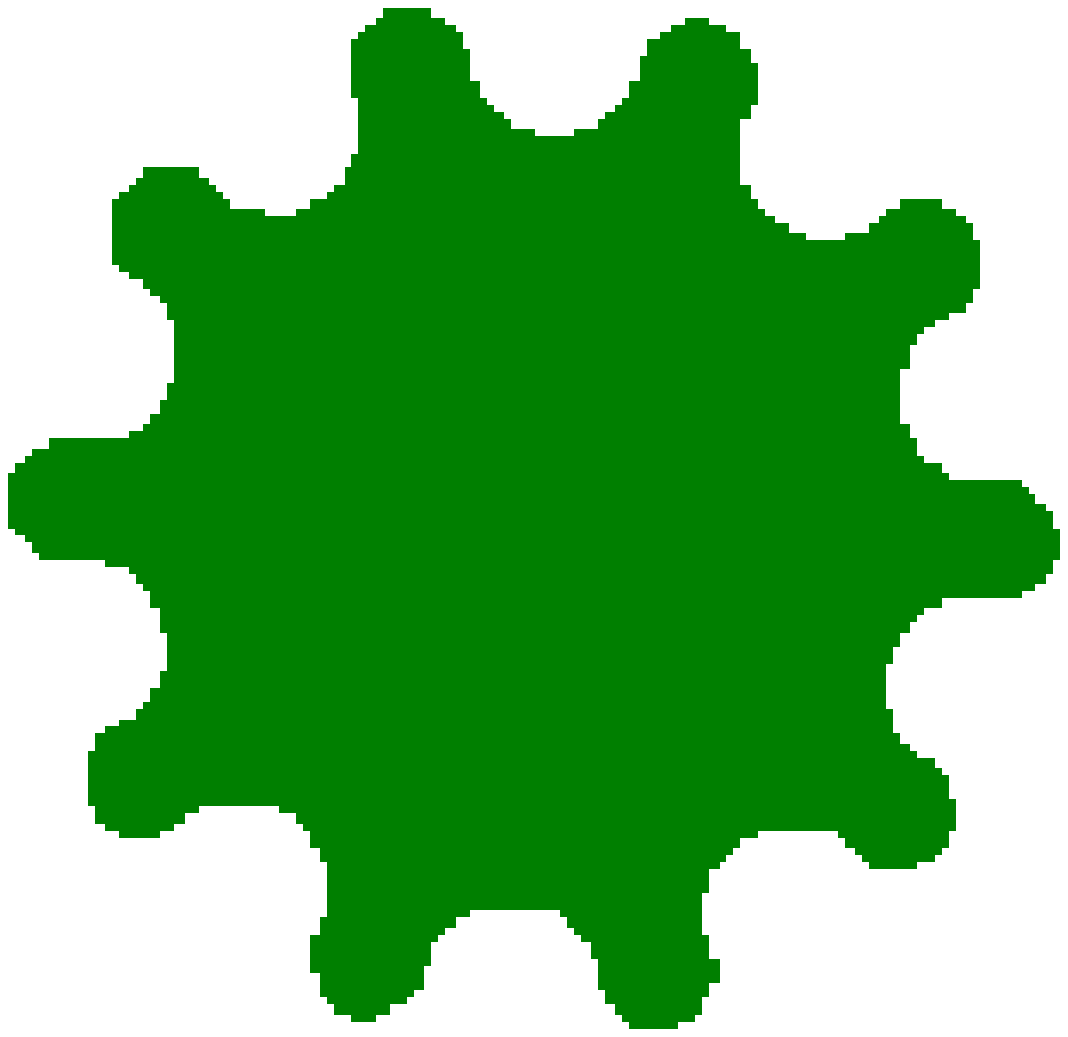}\hfill
\includegraphics[clip=true,width=0.16\textwidth]{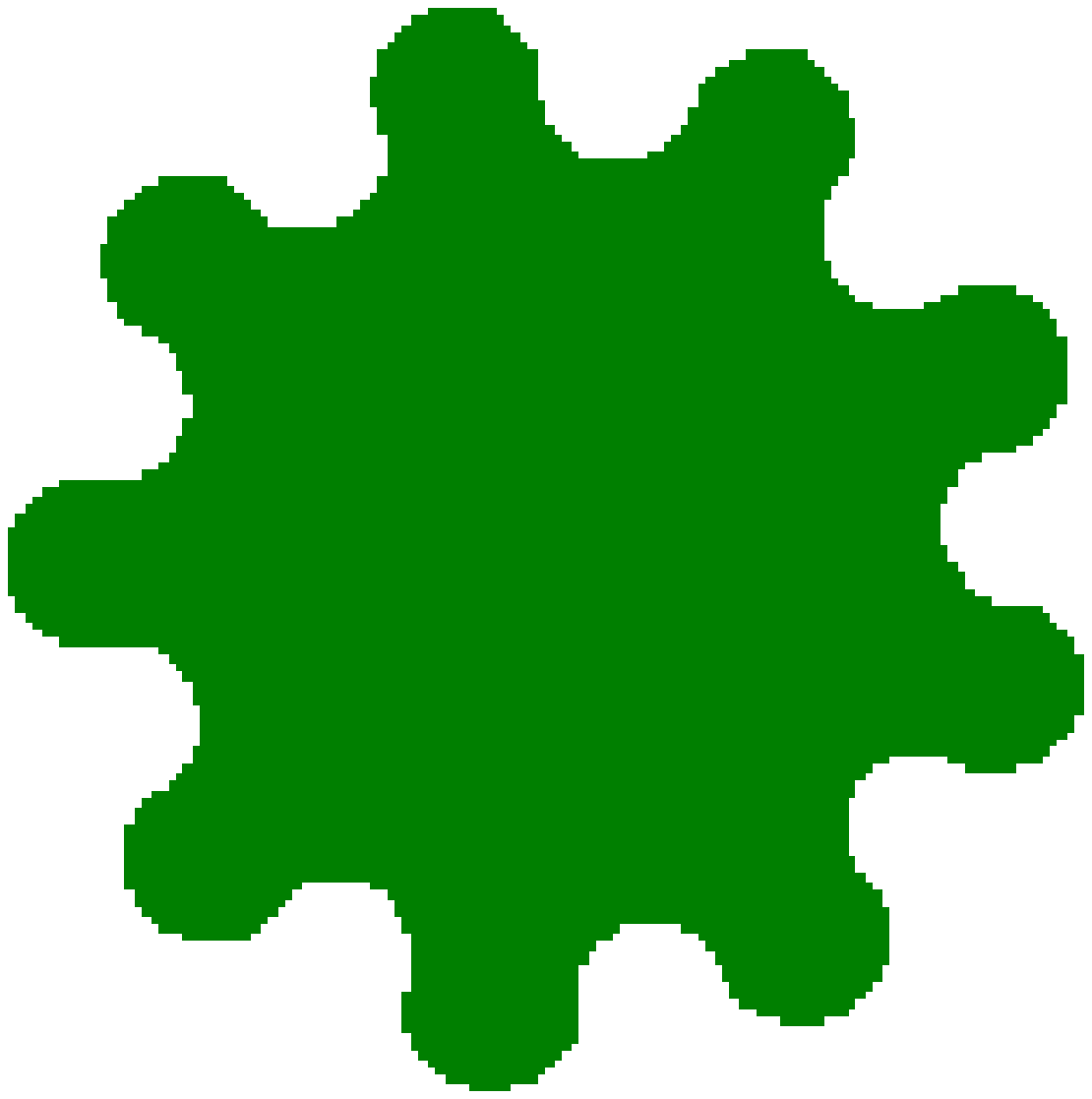}\hfill
\includegraphics[clip=true,width=0.16\textwidth]{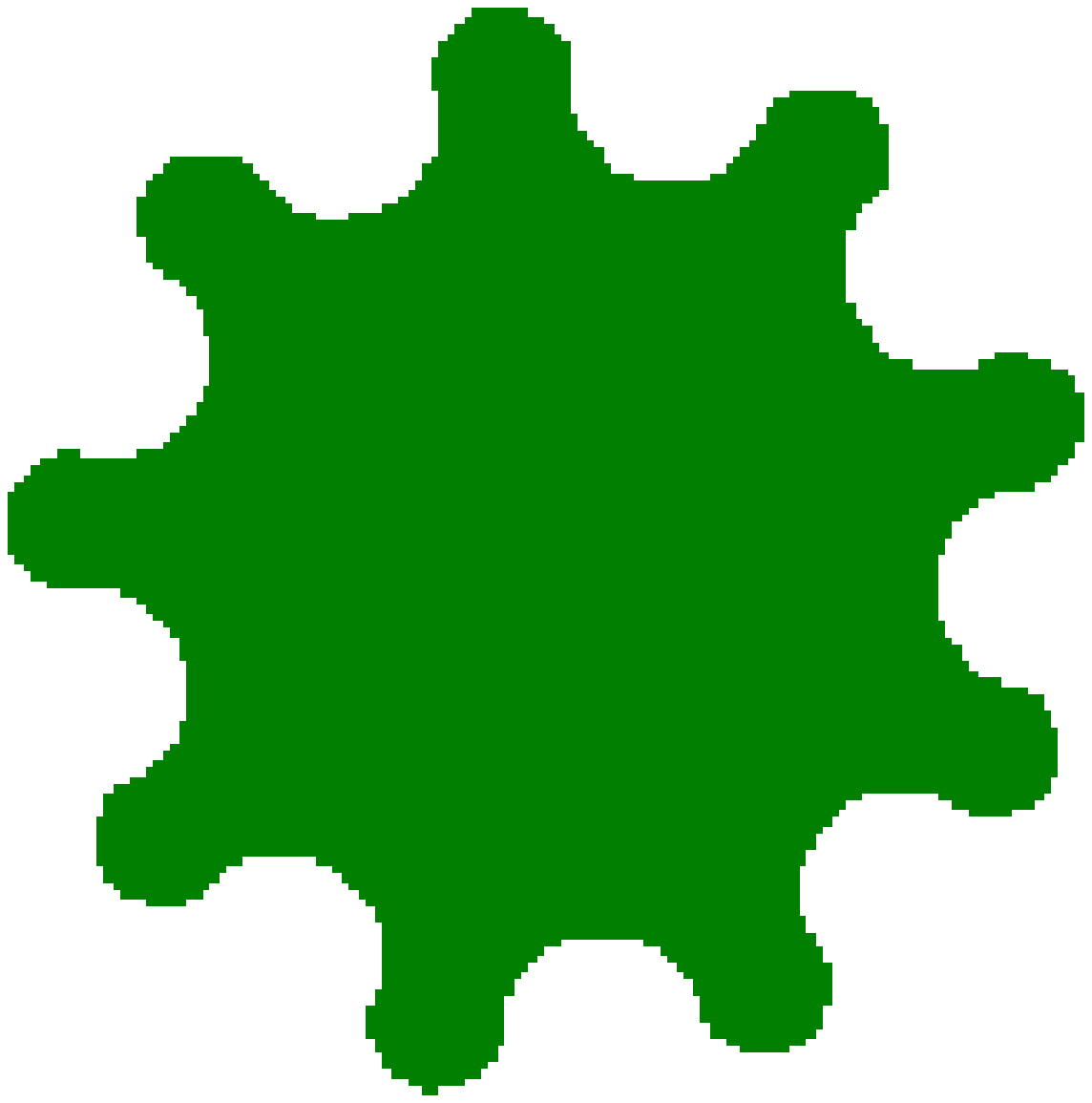}\hfill
\includegraphics[clip=true,width=0.16\textwidth]{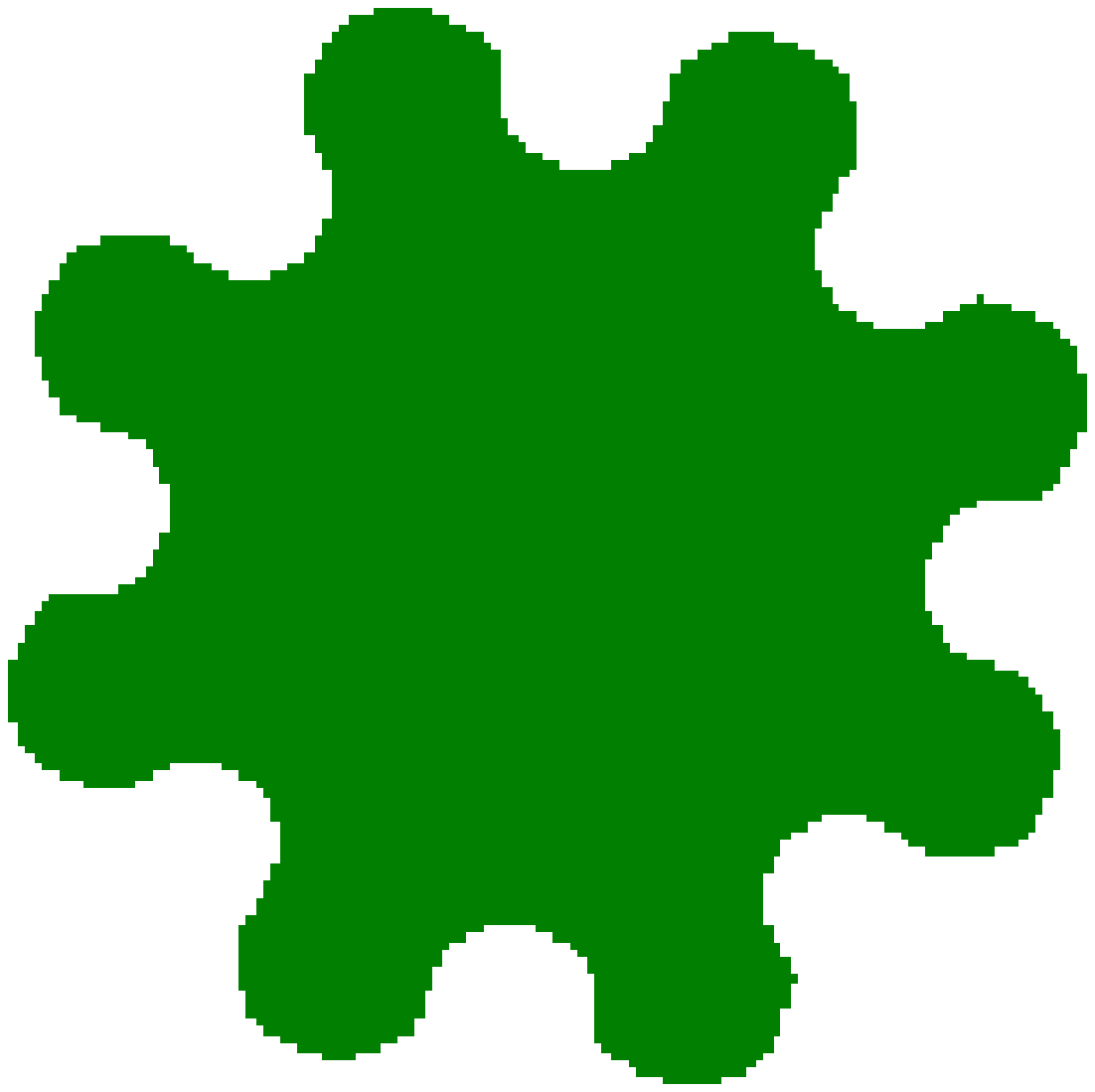}\hfill
\includegraphics[clip=true,width=0.16\textwidth]{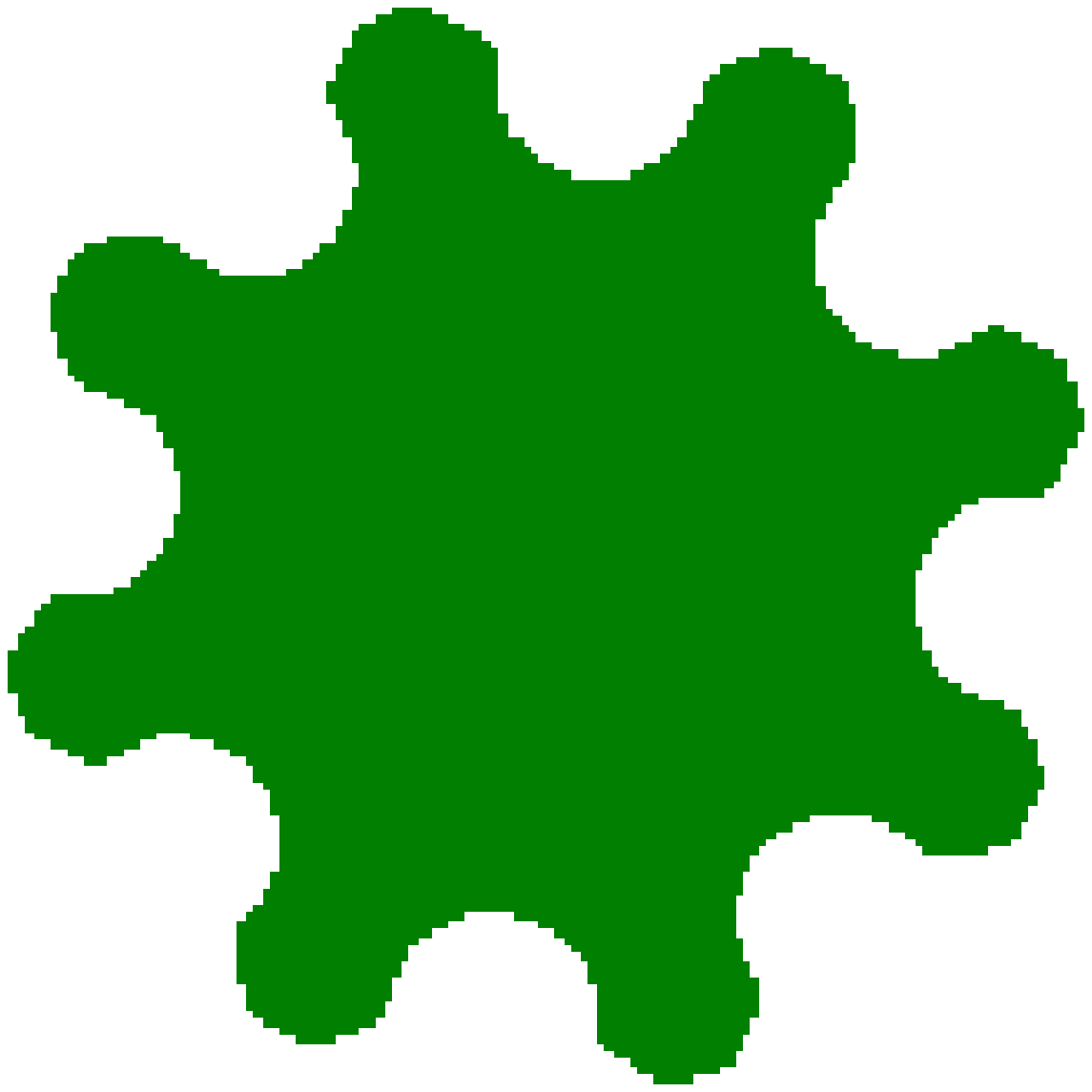}\hfill
\includegraphics[clip=true,width=0.16\textwidth]{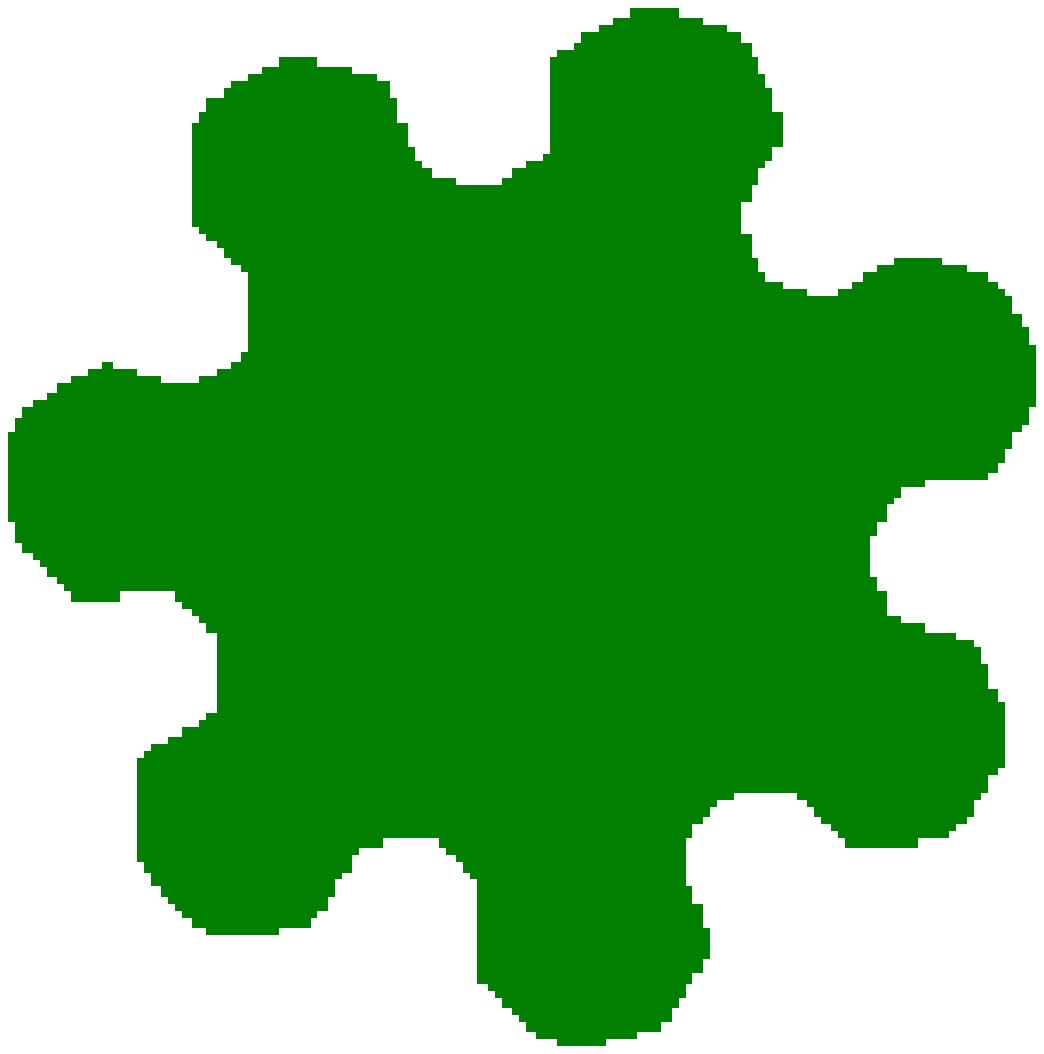}\hfill
\includegraphics[clip=true,width=0.16\textwidth]{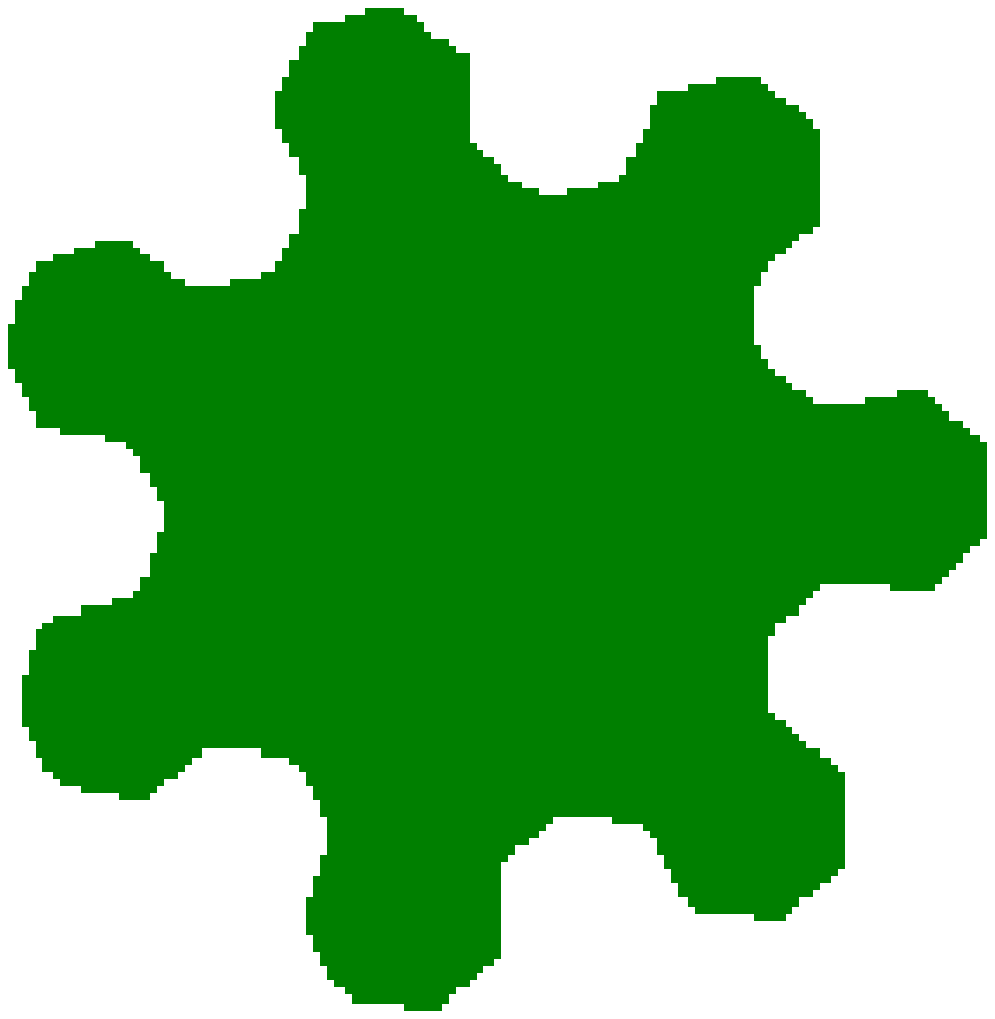}\hfill
\includegraphics[clip=true,width=0.16\textwidth]{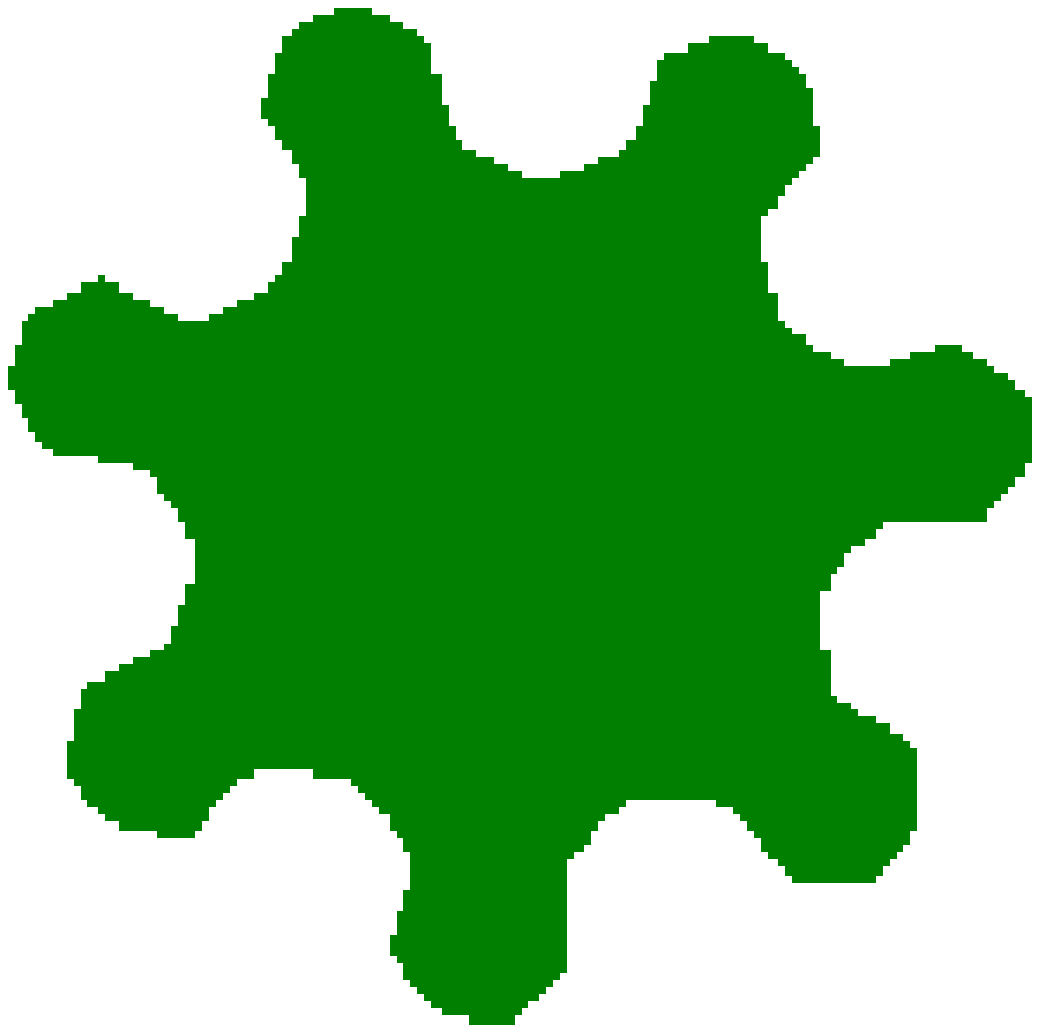}\hfill
\includegraphics[clip=true,width=0.16\textwidth]{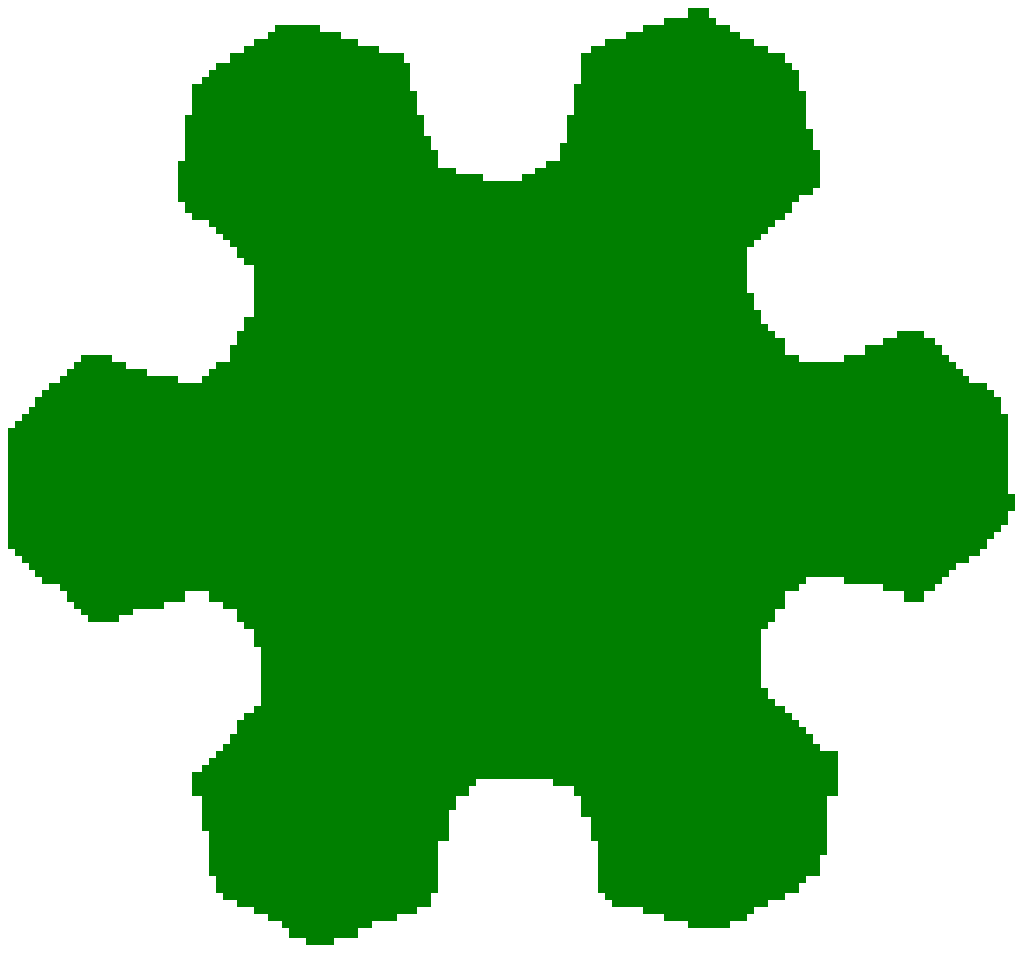}\hfill
\includegraphics[clip=true,width=0.16\textwidth]{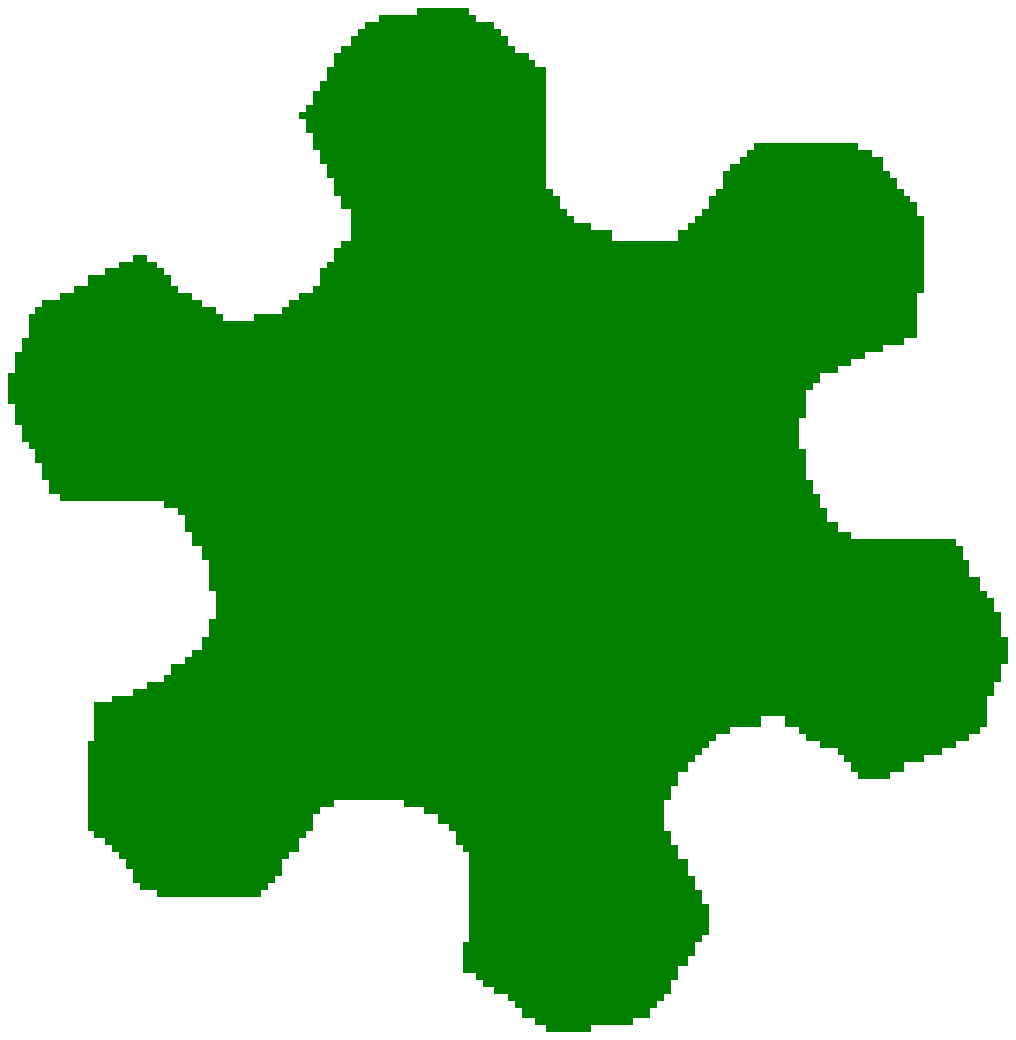}\hfill
\includegraphics[clip=true,width=0.16\textwidth]{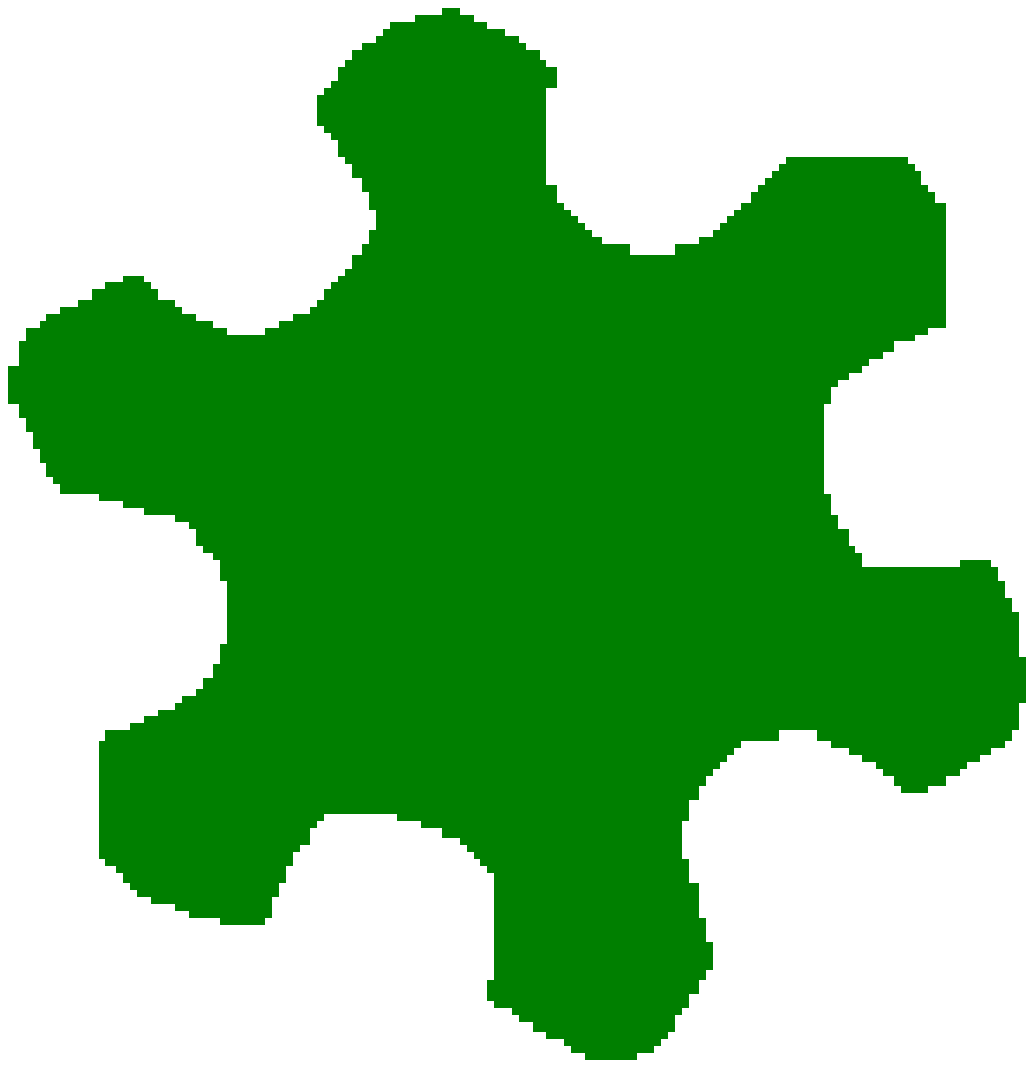}
\caption{\label{fig:cogs}Best configurations found for $d$ ranging
  from $0.22$ (top left) to $0.56$ (bottom right), where $d$ increases
  by $0.02$ in each subsequent panel going first from left to right
  and then from top to bottom.}
\end{figure}
We first consider $d<\pi^{-1/2}\approx0.564$. Figure~\ref{fig:cogs}
shows the best configurations found for $d$ between $0.22$ and $0.56$. These configurations have dihedral symmetry with a shape
resembling a cogwheel.  In particular, the cogwheel was always found superior to the
disc shape and was produced even if the starting configuration was
initialized in the shape of a unit disc (at sufficiently low
temperature) rather than a random distribution at high
temperature. This is consistent with our analytical results presented
in Sec.~\ref{sec:analytical}.

\begin{figure}
\includegraphics[clip=true,width=0.22\textwidth]{FiguresForPaper/bestconf028.eps}\hspace{15mm}
\includegraphics[clip=true,width=0.22\textwidth]{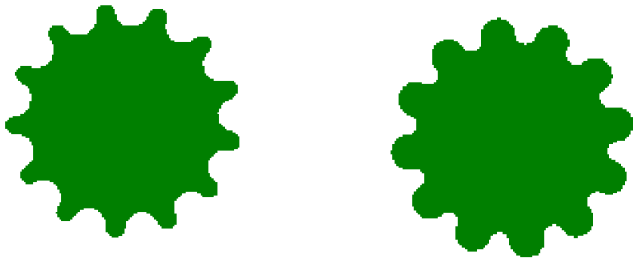}
\caption{\label{fig:twocogconfigs}At $d=0.28$ the configuration
  with 13 cogs (left) and the configuration with 12 cogs (right) have 
nearly the same value of $\discretep{s}{d}$.}
\end{figure}
\begin{figure}
\includegraphics[width=0.5\textwidth]{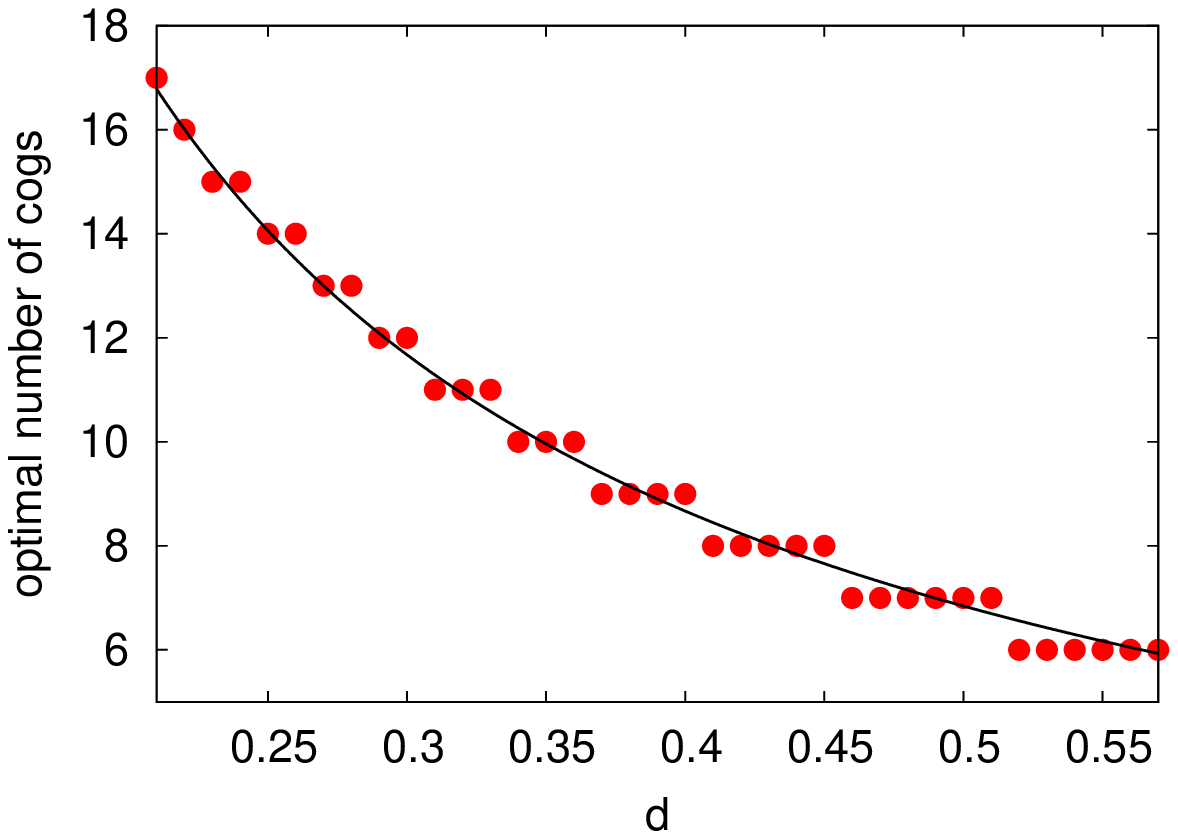}\hfill
\includegraphics[width=0.5\textwidth]{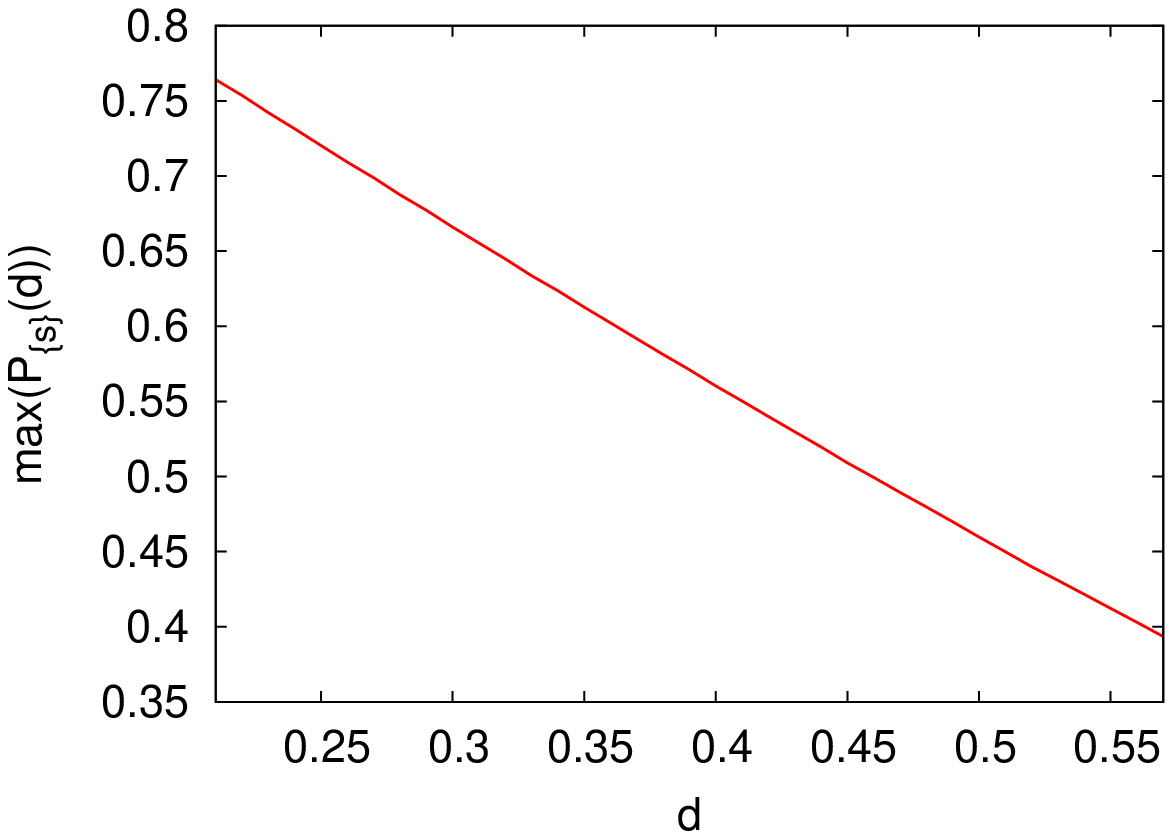}
\caption{\label{fig:optcogsandenergies}Left panel: optimal
  number of cogs according to our simulations (red dots). The number of cogs is well described by
  Eq.~(\ref{eq:nfoldsymm}) (solid line). Right panel: $\discretep{s}{d}$ for the
  best configurations found.}
\end{figure}

The optimal number of cogs, $n$, was found to decrease with increasing $d$ and is
well-described by the relation (\ref{eq:nfoldsymm}). However, the
number of cogs is discrete while $d$ can take any real value. The
optimal $n$ closely corresponds to the nearest integer satisfying
(\ref{eq:nfoldsymm}). If the solution of (\ref{eq:nfoldsymm}) is close
to a half-integer, the optimal number of cogs is difficult to 
identify precisely from the simulations: the best cogwheels found with either of the
two closest integer solutions to (\ref{eq:nfoldsymm}) yield very similar values of $\discretep{s}{d}$.
An example of two configurations with nearly the same $\discretep{s}{d}$ but with different
numbers of cogs is shown in Fig.~\ref{fig:twocogconfigs} for
$d=0.28$. The exact solution of (\ref{eq:nfoldsymm}) is $n=12.528\ldots$ in
this case, and hence $n= 12$ and $n=13$ are almost
equally good approximations. Figure~\ref{fig:optcogsandenergies} shows
the optimal number of cogs in the simulations as a function
of $d$, and the highest $\discretep{s}{d}$ found. 

\begin{figure}
\includegraphics[width=0.5\textwidth]{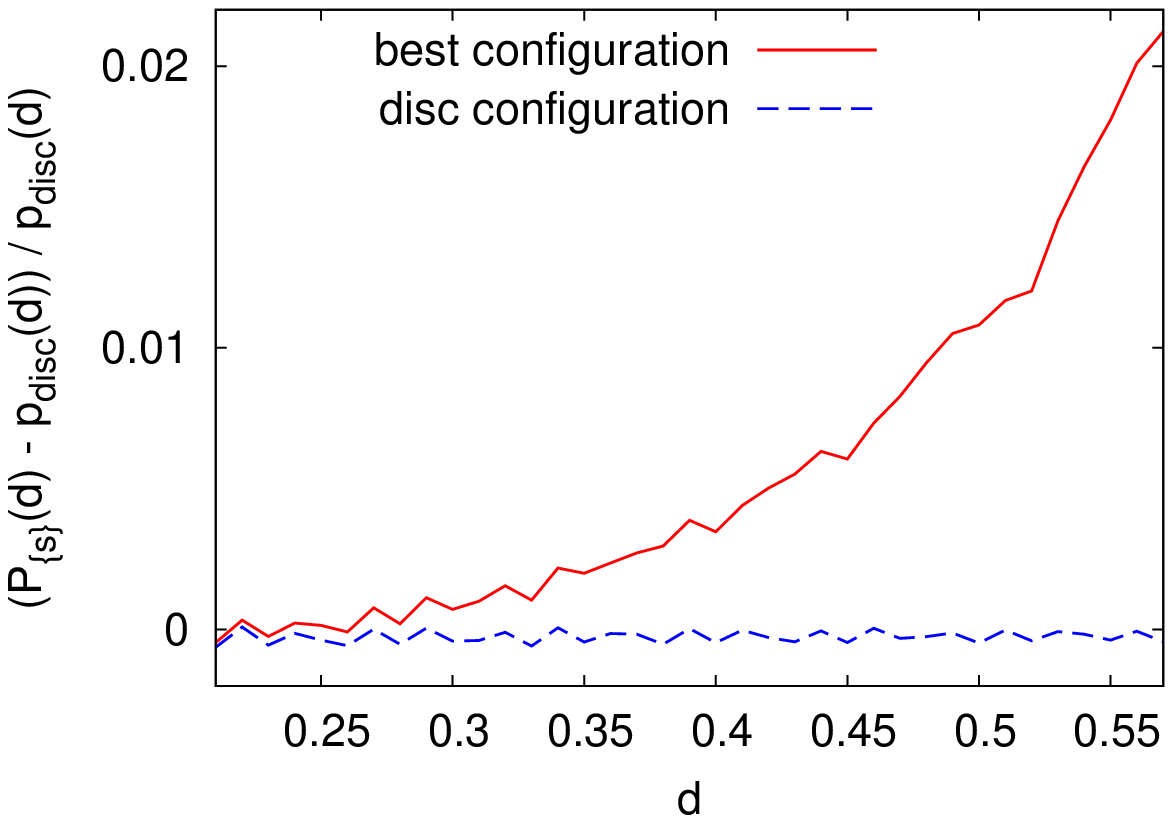}\hfill
\includegraphics[width=0.5\textwidth]{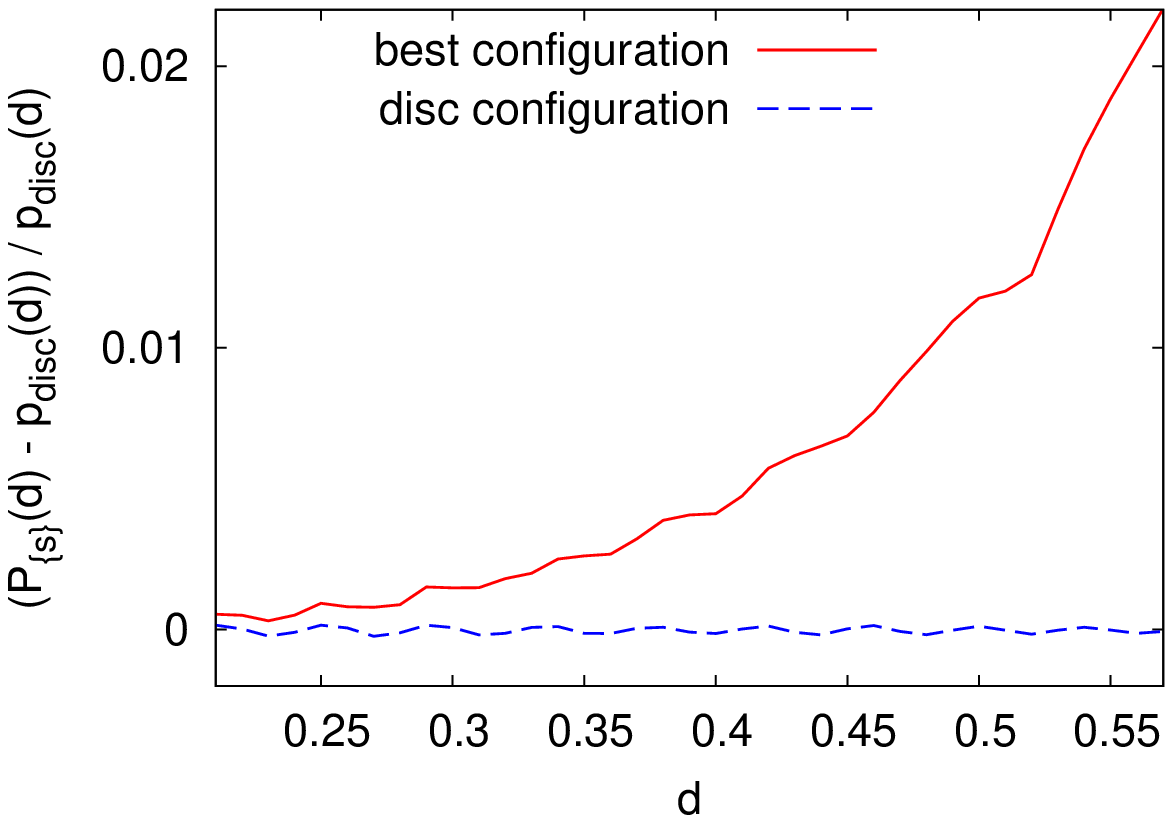}
\caption{\label{fig:disccogcomparison} Relative deviation of $\discretep{s}{d}$
from $\continuousp{\rm disc}{d}$ for the best cogwheel configurations
found (red solid line) and for the disc configuration (blue dashed
line) as a function of $d$. The discretization related fluctuations
are strongly correlated for the two types of configuration, and the
cogwheel was always found to be superior to the disc. Results shown
were obtained in the default setup with $N=10000$ (left panel) and
$N=90000$ (right panel). At larger $N$ fluctuations become milder and
the difference between the cogwheel and the disc configurations
becomes more pronounced.}
\end{figure}

\begin{figure}
\includegraphics[width=0.5\textwidth]{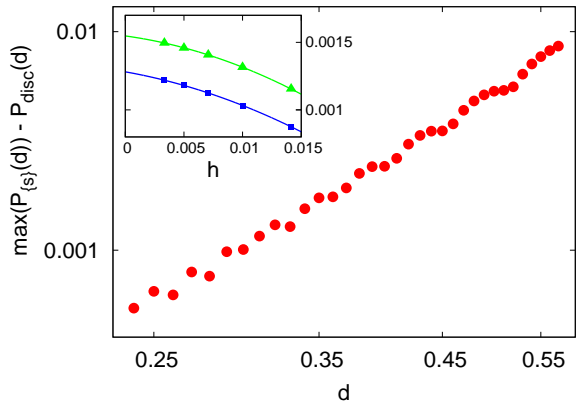}
\caption{\label{fig:disccogpowerlaw} Double-logarithmic plot of the
  difference between the largest value of $\discretep{s}{d}$ found,
corresponding to a
  cogwheel configuration, and $\discretep{s}{d}$ for the disc configuration, in the
  continuum limit $h\rightarrow0$. Inset: examples of (quadratic)
  continuum limit extrapolations for $d=0.33$ (blue squares) and
  $d=0.34$ (green triangles).}
\end{figure}

The results remained consistent for all setups and parameter values
studied. Moreover, the orientation of the cogwheel configurations with
respect to the grid was different for independent simulation runs and
appeared uncorrelated with the symmetry axes of the grid. This was the
case also for the configurations at higher $d$ discussed in the
following sections. At low $d$, the difference in the values of $\discretep{s}{d}$
for the best cogwheel configurations found and the disc shaped
configurations was very small. In particular, it was sometimes found
to be smaller than the fluctuations due to the discretization that are
shown in Fig.~\ref{fig:consistencychecks}. These fluctuations,
however, appear to be strongly correlated for the two types of
configuration, and the cogwheel was found to be consistently superior
to the disc for all $d$ and in all discretization setups considered,
see Fig.~\ref{fig:disccogcomparison}. At larger $N$ fluctuations
become milder and the difference between the cogwheel and the disc
configurations becomes more pronounced.
Figure~\ref{fig:disccogpowerlaw} shows the difference between the
values of $\discretep{s}{d}$ for the best cogwheel and the disc configurations,
plotted as a function of $d$ on a double-logarithmic scale. This
difference was calculated independently for several lattice spacings
and then extrapolated to the continuum limit $h \rightarrow 0$. It is
always positive and appears to be roughly a straight line on large
scale, modulated on smaller scales by fluctuations associated with the
relationship of $d$ to the number of cogs.   The approximate linearity
suggests a possible power law. 
The continuum limit $h\rightarrow0$ was
taken by quadratically extrapolating in the lattice spacing, which is
given by $h=1/\sqrt{N}$ for a square lattice. Examples of such
continuum extrapolations are shown in the inset of
Fig.~\ref{fig:disccogpowerlaw}. 

\begin{figure}[ht]
\includegraphics[width=0.24\textwidth]{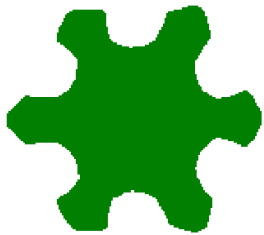}
\vspace{-6mm}
\caption{\label{fig:d057}Best configuration found for $d=0.57$.}
\end{figure}
A cogwheel was also the best configuration found for $d=0.57$, slightly above
$\pi^{-1/2}$. In this case, shown in Fig.~\ref{fig:d057}, a slight
symmetry breaking from $D_6$ to $D_3$ dihedral symmetry in the shapes
of the cogs appears to develop. As we
argued in Sec~\ref{sec:analytical}, there is some reason to expect
that the optimal lawn shape may fail to be simply connected for
$d\gtrsim\pi^{-1/2}$.  Our simulations suggest this is indeed the
case.  Indeed, the best configurations found for $d\geq0.58$ are all 
not only not simply connected but disconnected. They will be
discussed in the following sections.  It is interesting
to observe this abrupt transition from connected optimal lawns to
disconnected ones, which our results suggest occurs
for $d$ between $0.57$ and $0.58$.   Finer scale discretizations
should allow a closer exploration of the transition regime
and its (dis)analogies with other transitions in statistical 
models.

\subsection{Critical regime}
\begin{figure}[ht]
\includegraphics[width=0.24\textwidth]{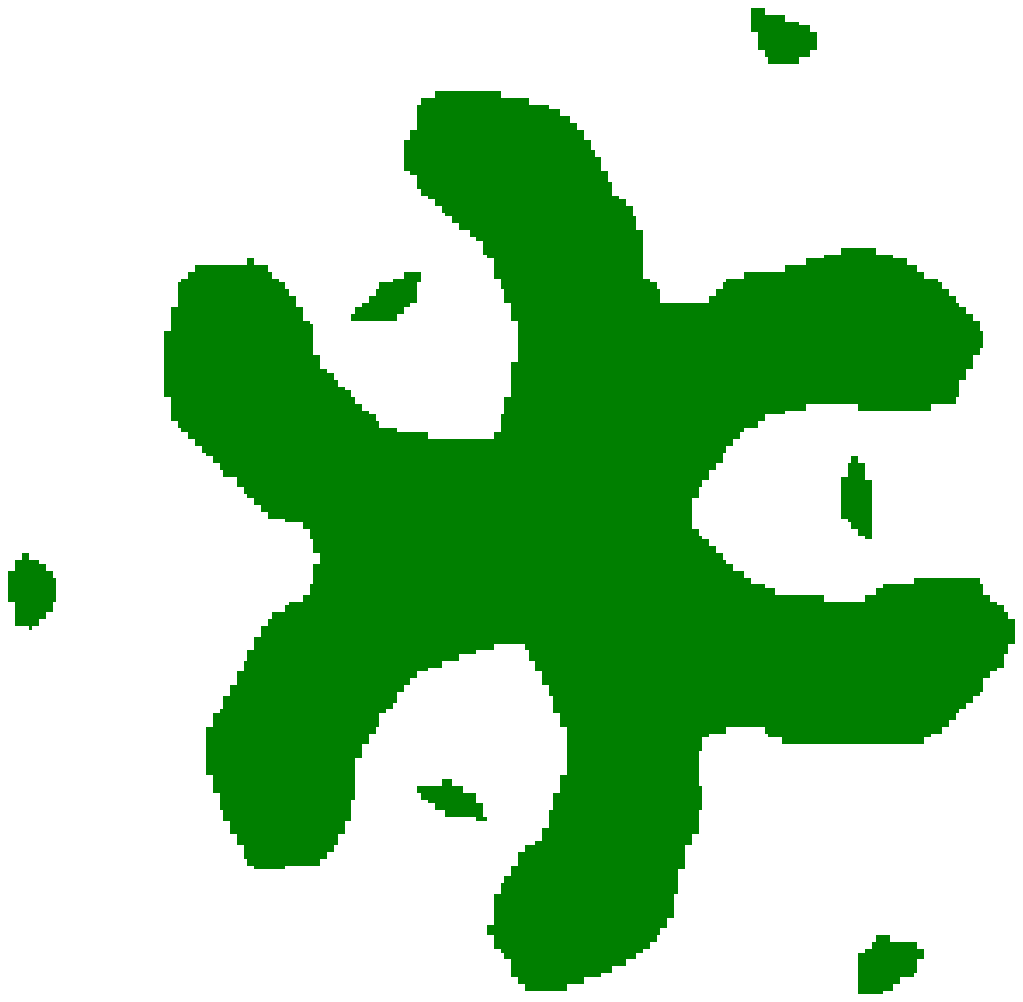}\hfill
\includegraphics[width=0.28\textwidth]{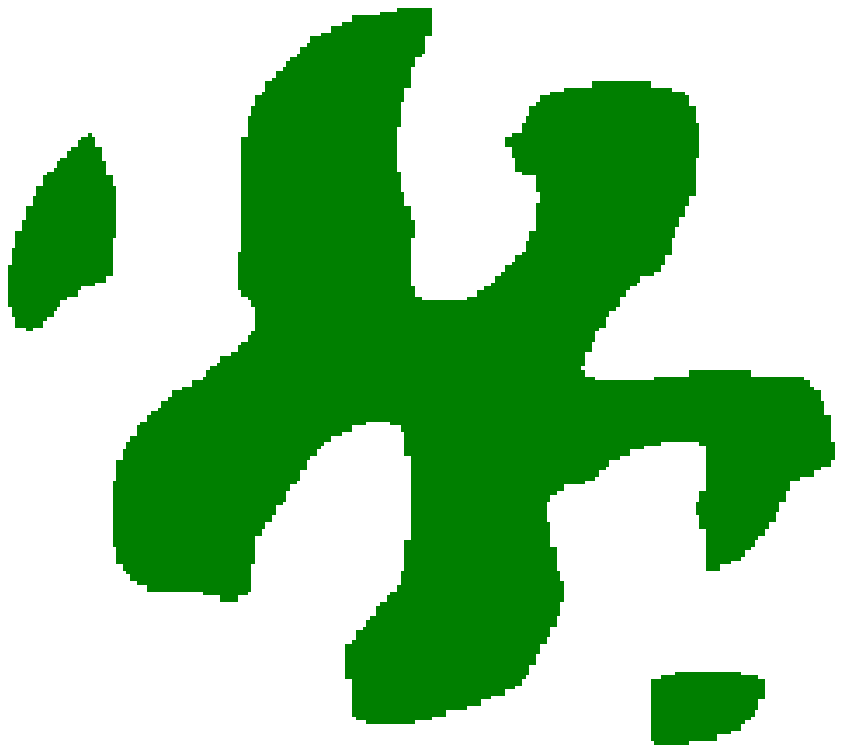}\hfill
\includegraphics[width=0.24\textwidth]{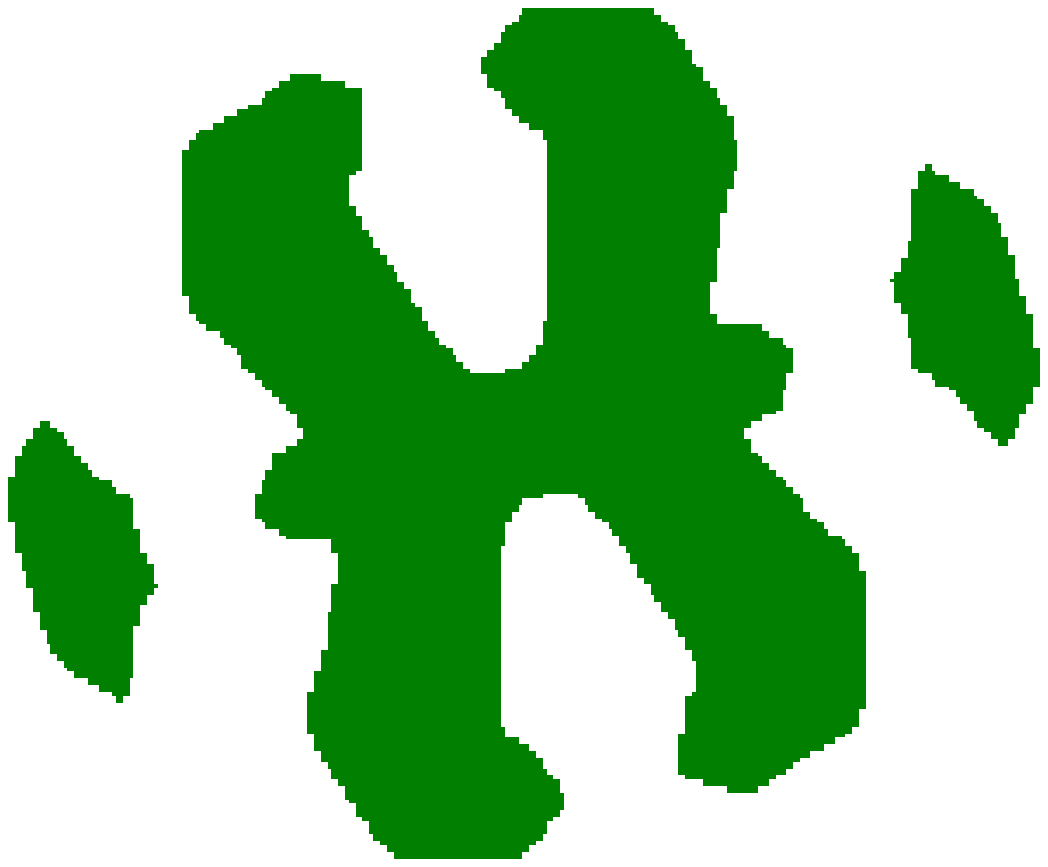}
\vspace{-6mm}
\caption{\label{fig:critical}Three types of shape that were found to
  be optimal for $0.58\leq d\leq0.64$. All shapes are
  disconnected. Left panel: Best configuration found for $d=0.58$ (the
  same type of shape was also the best found for $d=0.59$). Middle
  panel: Best configuration found for $d=0.60$. Its shape appears to
  lack any symmetry. The same type of shape was the best found for
  $d=0.61$ (in most discretization setups) and found to be nearly
  optimal for a range of $d$ between 0.58 and 0.61. Right panel: Best
configuration found for $d=0.62$. The same type of shape was the best
found for $0.62\leq d\leq0.64$ and found to be nearly optimal for a
range of $d$ between 0.59 and 0.66.}
\end{figure}

For $d\geq0.58$ the best configurations found by the simulations
  exhibit unexpectedly complex disconnected shapes. In the interval
  $0.58\leq d\leq0.64$ we found three distinct types of shape,
  which are shown in Fig.~\ref{fig:critical}. Each shape appears
  optimal for only a small range of $d$ within this interval. The
  simulations also frequently generated suboptimal configurations
  corresponding to local maxima (such
  configurations will be discussed in Sec.~\ref{sec:numsummary}).

For $0.58\leq d\leq0.59$ the best shape found has three-fold rotational
and mirror symmetries (see left panel of
Fig.~\ref{fig:critical}), corresponding to the dihedral group $D_3$. Despite being superior to other types of
configuration generated in this range of $d$, the three-fold shape was
only infrequently found by the simulations, suggesting that it is
located at a maximum that is difficult to reach. The best shape
found for $0.60\leq d\lesssim0.61$ appears to lack any symmetry. It is
shown in the middle panel of Fig.~\ref{fig:critical}. Close to
$d=0.61$ a transition to a roughly ``H''-shaped solution with
additional disconnected patches was observed. This shape, shown in the
right panel of Fig.~\ref{fig:critical} is mirror symmetric around two orthogonal
axes, corresponding to the dihedral group $D_2$. Configurations of this type were the best found for
$0.62\leq d\leq0.64$. At $d=0.61$ the generated H-shaped
configurations had nearly the same $\discretep{s}{d}$ as the asymmetric ones,
suggesting that the transition between the asymmetric and the H-shaped
regimes occurs close to $d=0.61$.  H-shaped configurations were also
occasionally generated as (apparently) nearly optimal configurations for values of
$d$ between 0.59 and 0.66, while the asymmetric configurations were
frequently generated for $d$ between 0.58 and 0.61. As opposed to the
three-fold shape, these two types of configuration appear to be
located in maxima of the model that are easier to reach. The abundance of
different shapes generated in this regime and the rapid transitions
between them as $d$ is varied suggest that this regime has a
particularly interesting and complex landscape. It
also makes it more plausible that not all maxima in this regime 
have been explored in the present setup and better configurations
might yet be discovered.

\subsection{Three-bladed fan regime}
\begin{figure}[ht]
\hspace{-2mm}\includegraphics[width=0.2\textwidth]{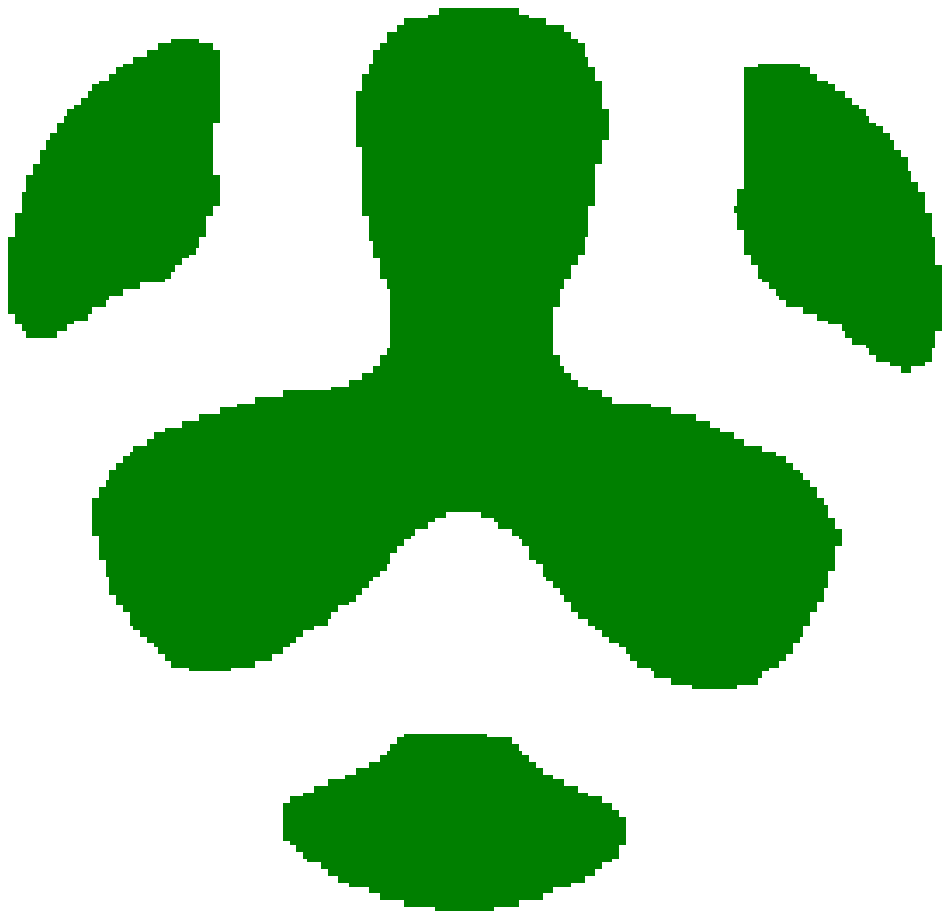}\hfill
\includegraphics[clip=true,width=0.2\textwidth]{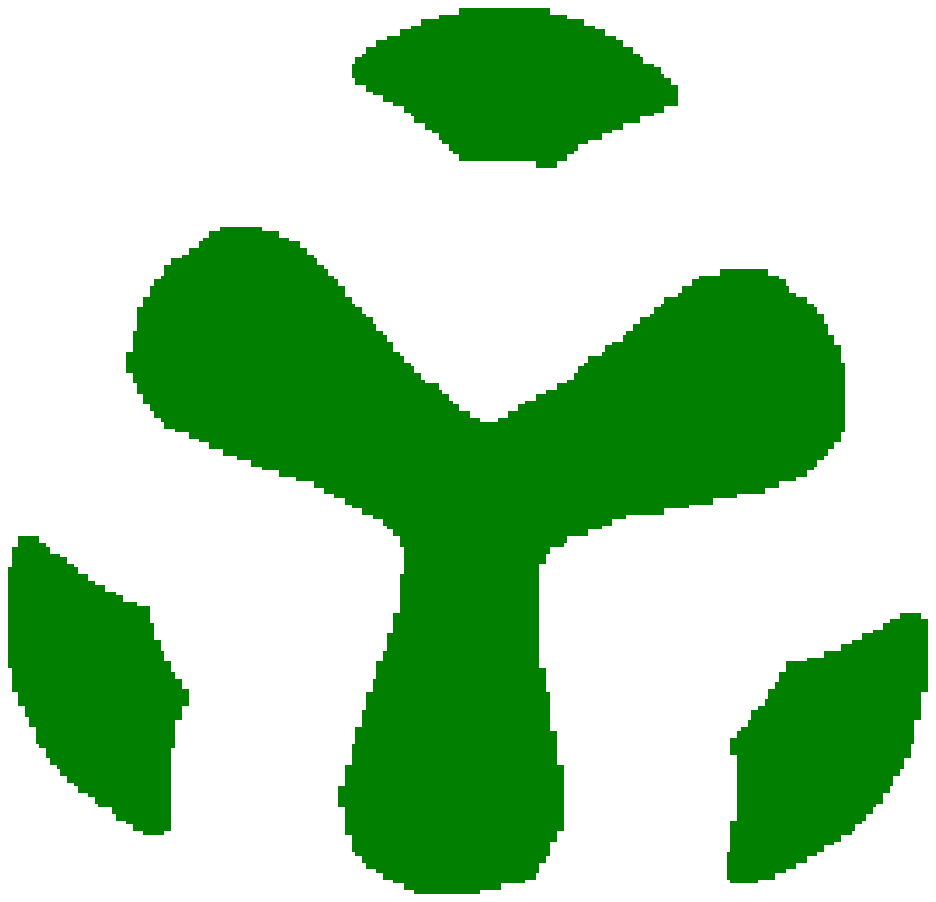}\hfill
\includegraphics[clip=true,width=0.2\textwidth]{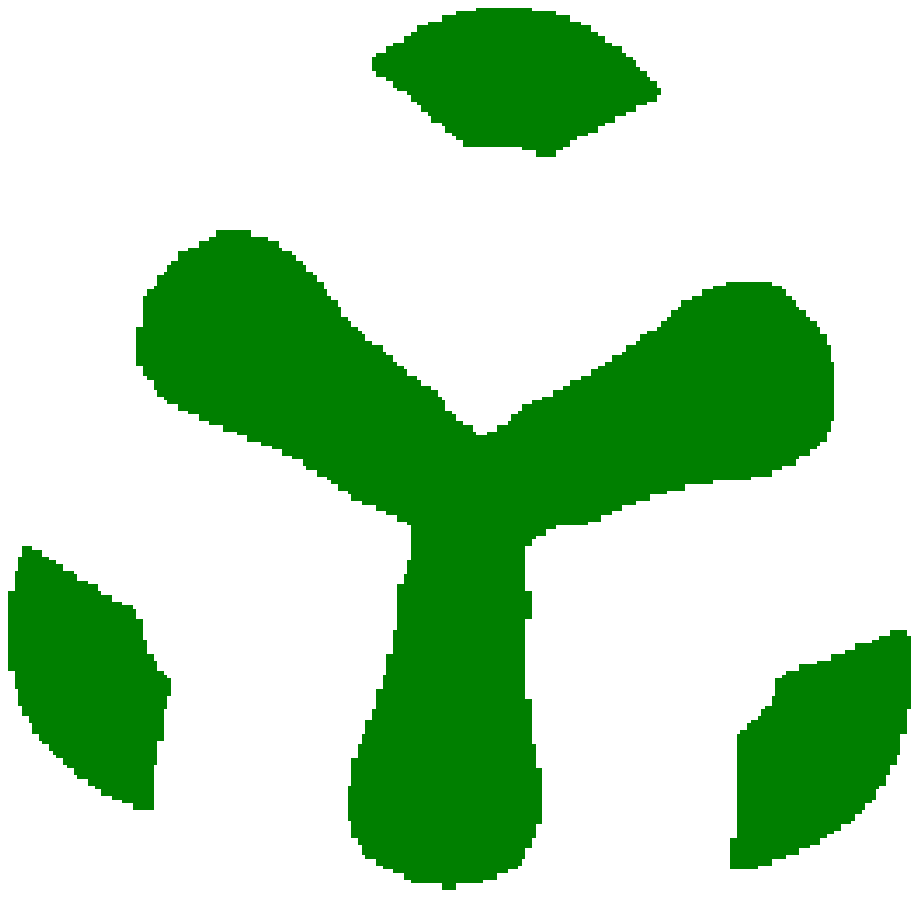}\hfill
\includegraphics[clip=true,width=0.2\textwidth]{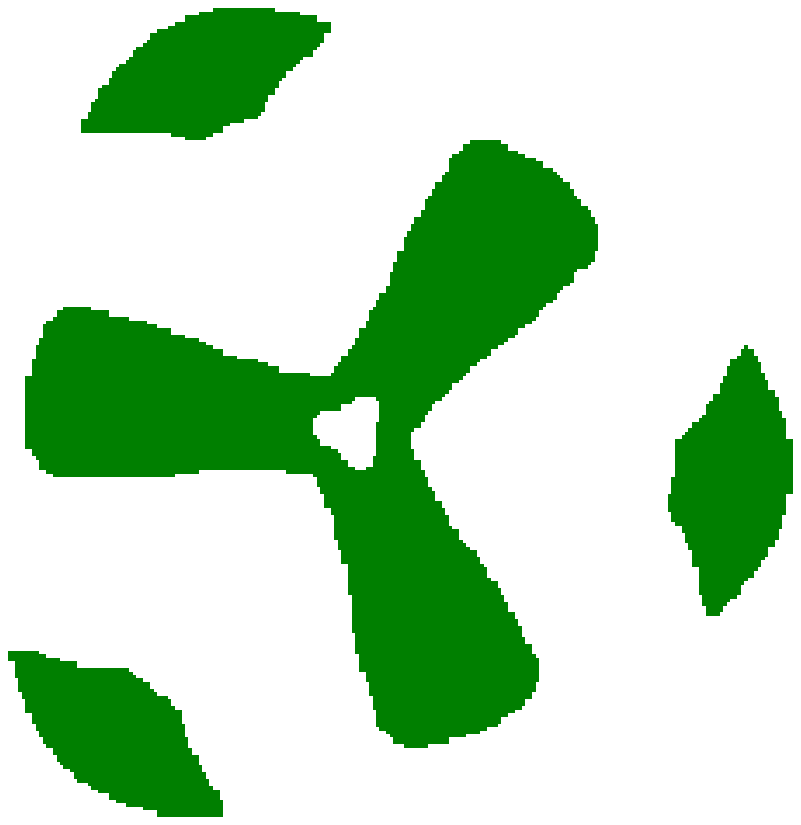}\hfill
\includegraphics[clip=true,width=0.2\textwidth]{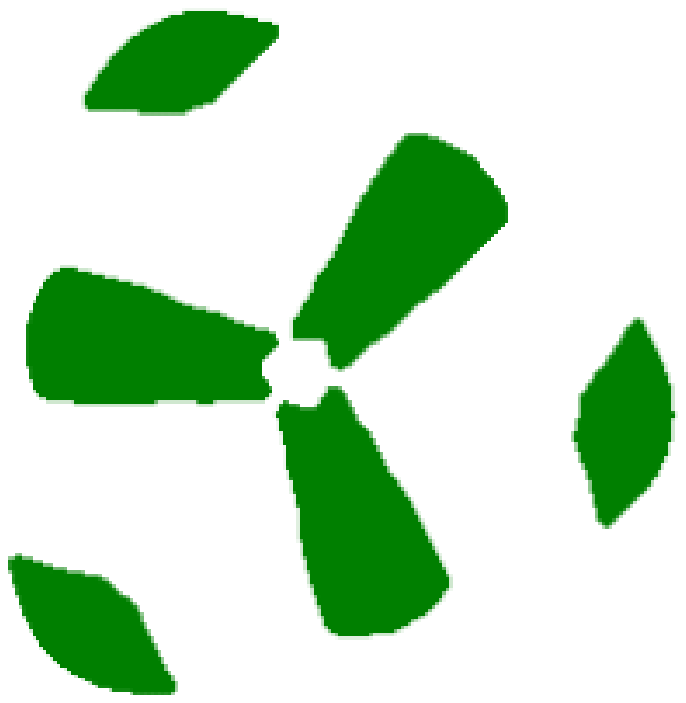}
\caption{\label{fig:3bladedfan}Best configurations found for $d=0.65$, 0.72, 0.78, 0.82, 0.87 (left to right).}
\end{figure}

For $0.65\leq d\leq0.87$ the best configurations found are shaped like
a three-bladed ``fan'' with additional patches between the blades, as
shown in Fig.~\ref{fig:3bladedfan}. These configurations have
three-fold rotational and mirror symmetries, corresponding to the dihedral group $D_3$. Starting from
approximately $d=0.78$ a
hole starts to develop in the centre of the fan. The hole becomes
larger with increasing $d$. The three-bladed fan was also occasionally
generated as an (apparently) nearly optimal configuration at larger values of $d$.

\subsection{Stripes regime}
As $d$ is further increased, the best configurations found by the
simulations comprise four roughly parallel ``stripes'' of finite
length. The transition occurs close to $d=0.88$ where the values of $\discretep{s}{d}$ for
the fan and the stripes configurations are comparable. Configurations
with five stripes were also sometimes generated, but had lower
$\discretep{s}{d}$. The distance between the stripes roughly corresponds to $d$
and their boundaries consist of several curved segments. Stripes that
are further away from the centre are thinner and shorter than the
central ones. The configuration with four stripes appears related to
the H-shaped configuration, with the connecting part missing and the
disconnected patches further apart. Like the H-shaped configuration,
the stripes configurations is mirror symmetric around two orthogonal
axes, corresponding to the dihedral group $D_2$.

\begin{figure}[ht]
\includegraphics[clip=true,width=0.24\textwidth]{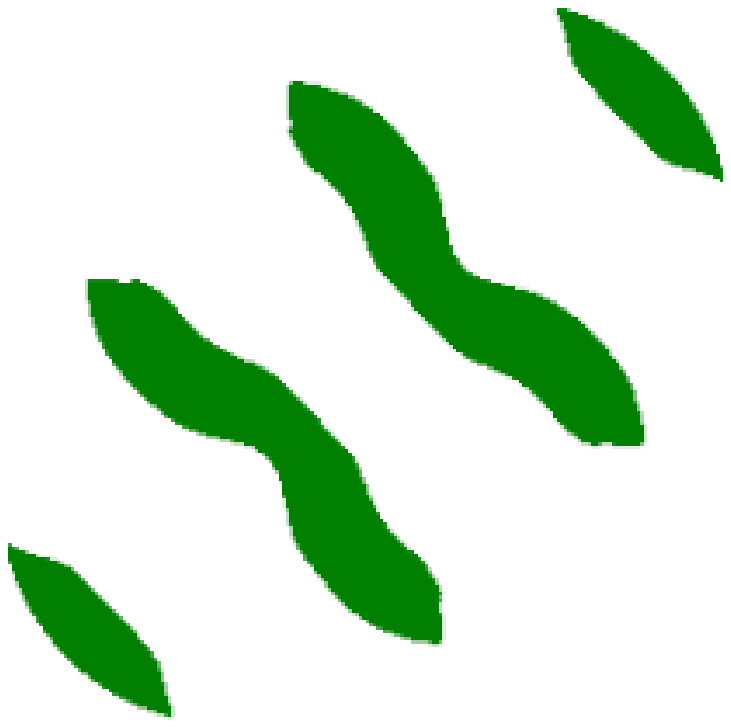}\hspace{15mm}
\includegraphics[clip=true,width=0.24\textwidth]{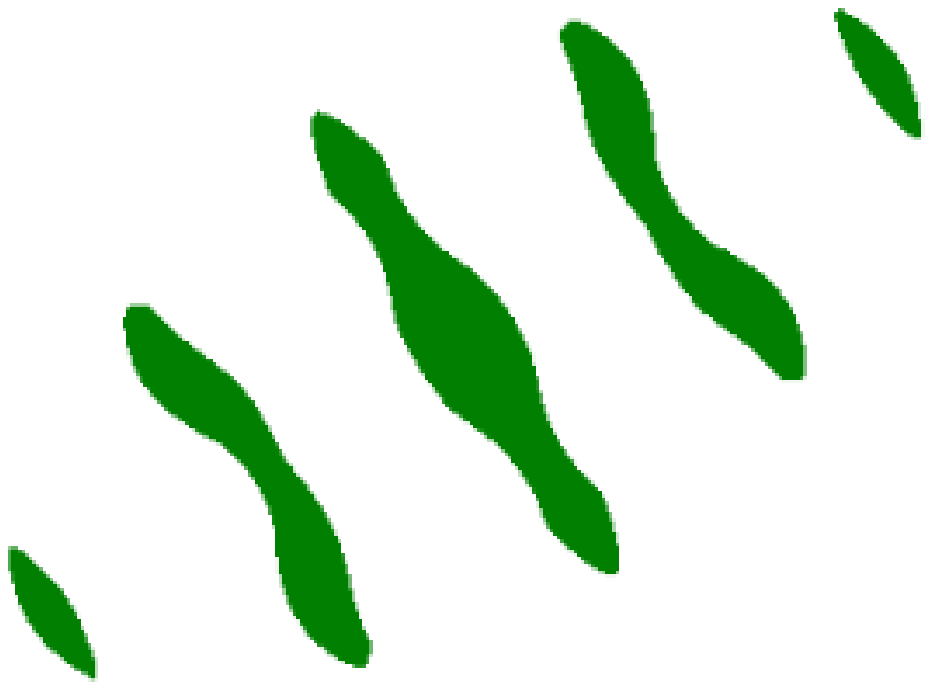}
\vspace{-6mm}
\caption{\label{fig:stripes}The best configuration found for $d=0.9$
  consists of four stripes (left). The same type of shape was the best
  found for $d\geq0.89$ and up to the largest $d$ considered in this
  study. The five striped configuration (right) has lower $\discretep{s}{d}$.}
\end{figure}

For the range $d\leq1.0$ considered in this work, no configuration
with more than five stripes was found. We do not know
whether configurations with more stripes will 
emerge for larger $d$.  Figure~\ref{fig:stripes} shows an example of a four
striped configuration found to be the best for $d=0.9$ and a five striped
configuration found to be nearly as good.

\subsection{Summary of numerical results}
\label{sec:numsummary}
\begin{figure}[ht]
\includegraphics[clip=true,width=0.99\textwidth]{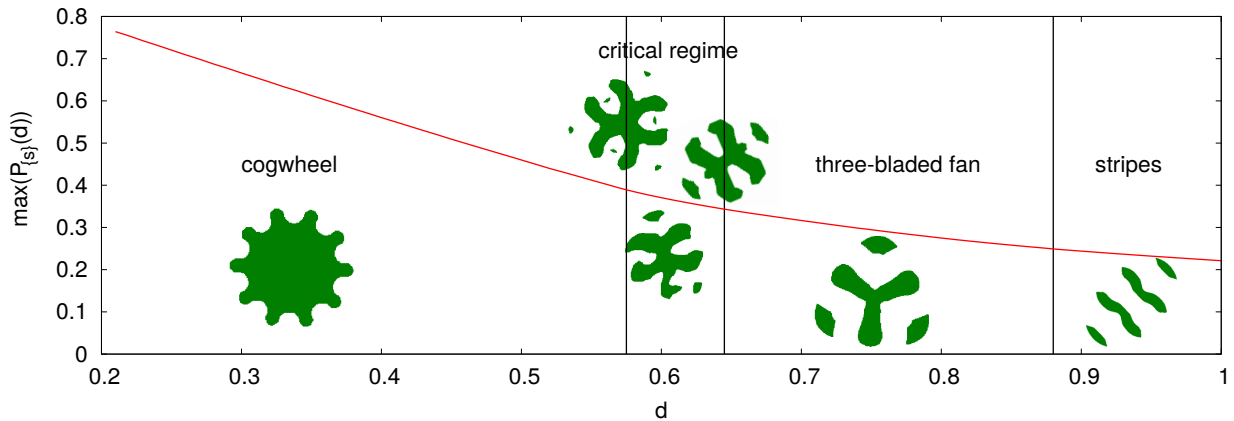}
\caption{\label{fig:phasediag}Diagram of the different regimes and the
  corresponding maximal values found for $\discretep{s}{d}$ as a function of $d$.}
\end{figure}
Figure~\ref{fig:phasediag} summarises the ``phase diagram''
of the different regimes with corresponding highest values of $\discretep{s}{d}$ obtained in the default setup.

\begin{figure}[ht]
\includegraphics[clip=true,width=0.22\textwidth]{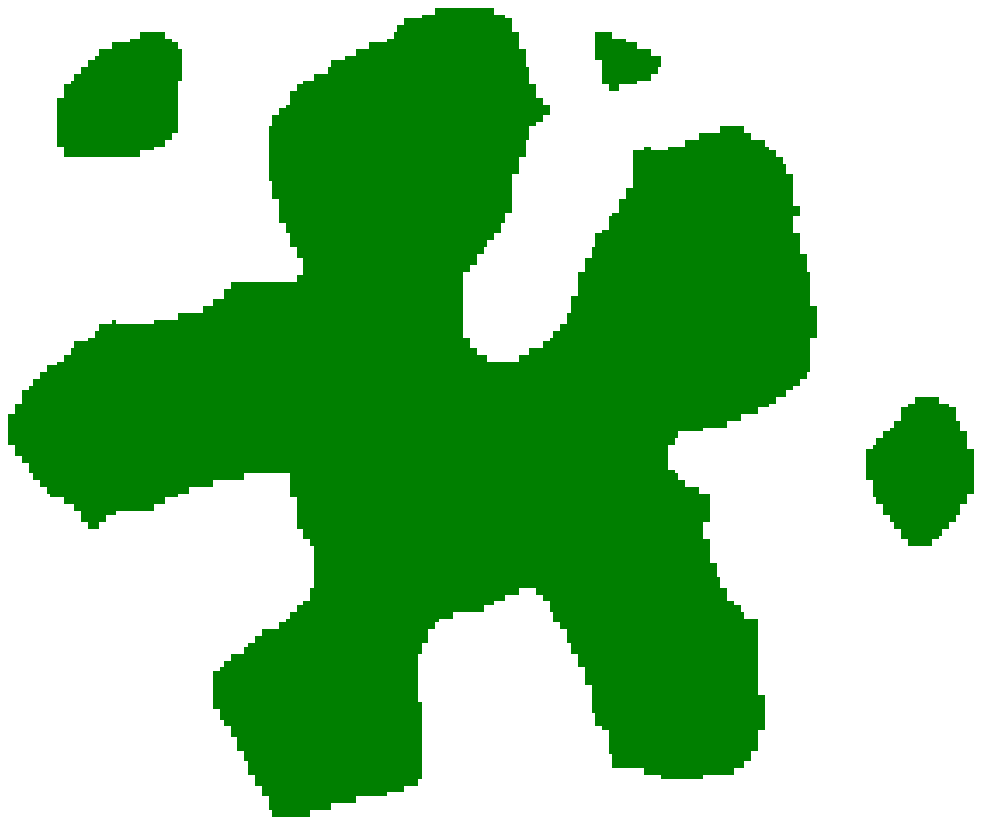}\hfill
\includegraphics[clip=true,width=0.22\textwidth]{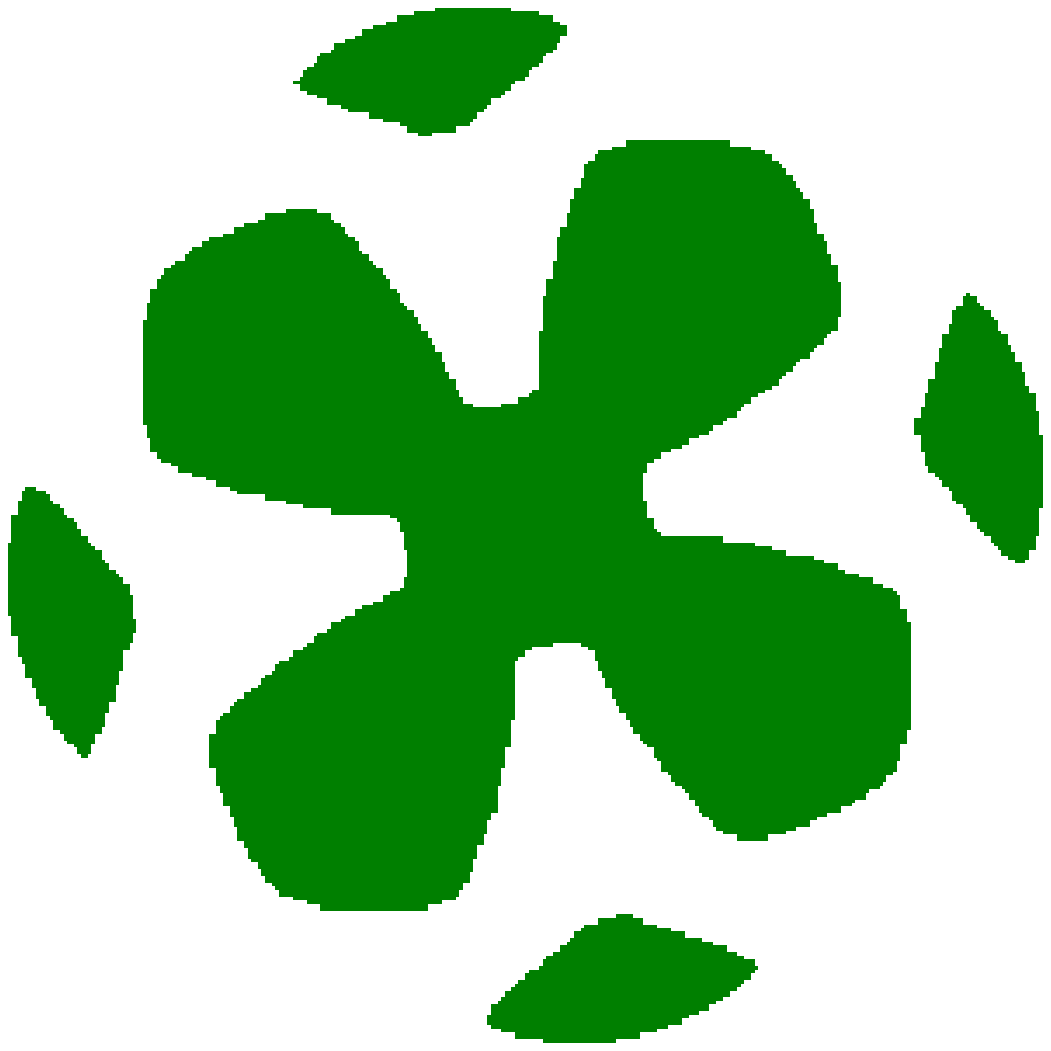}\hfill
\includegraphics[clip=true,width=0.22\textwidth]{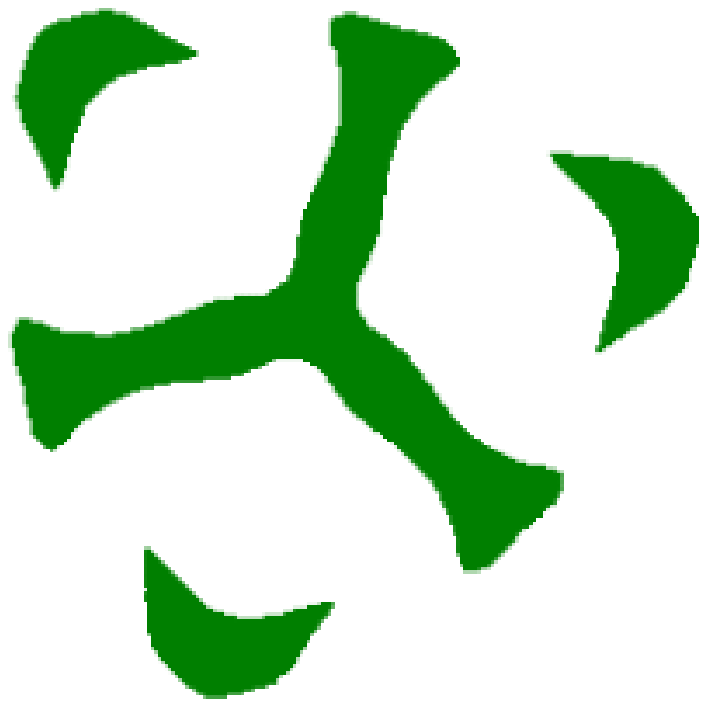}
\vspace{-3mm}
\caption{\label{fig:altconfs}Suboptimal configurations generated by
  the annealing process. Left panel: mirror-symmetric shape with five
  prongs generated for $d=0.58$ (critical regime). Middle panel:
  four-bladed fan configuration generated for $d=0.61$ (critical regime). Right panel: three-fold
  symmetric shape generated at $d=0.95$ (stripes regime).}
\end{figure}
As discussed in the previous sections, the system occasionally
annealed to suboptimal configurations, corresponding to local maxima
of $\discretep{s}{d}$. Usually these configurations were of a shape type that was
the best found for other values of $d$, but sometimes new shapes were also
generated. Examples of such shapes are shown in Fig.~\ref{fig:altconfs}. Of
course, it is possible that other near-optimal (or even
optimal) shapes exist which were
not found by the present simulations. In particular, shapes with
very fine structures that cannot be resolved with the present 
discretization would require larger values of $N$.

\section{Discussion and outlook}

In summary, we take our numerical results as strongly suggesting
a qualitative description of the broad features of solutions to the 
grasshopper problem for the ranges of $d$ investigated. 
This description remains consistent under various non-trivial consistency checks,
and is also consistent with our analytic results. 
We hope that further refining our discretization may allow 
a sufficiently precise description of the apparently optimal
configurations to allow the possibility of conjectures amenable
to analytical proof.   It should also allow a more precise
characterisation of the values of $d$ 
defining transitions between regimes that characterise qualitatively different
shapes, particularly those around the range $0.57 \leq d \leq 0.64$,
where the solution space appears to have a particularly complex structure.
 
Further numerical and analytic investigations are thus definitely
merited. It would also be worthwhile to study the behaviour of the
spin model itself in greater detail. The focus of this work is on
finding the ground state of the system. Other interesting
questions include the behaviour of the system at finite temperatures
(and possible associated transitions), as well as transitions
associated with varying lattice size and number of spins,
thermalization times, the calculation of thermodynamic observables,
and so on.

We would not necessarily expect our results for large $d$ to
extrapolate well to the grasshopper problem on the sphere when
the lawn has the area of the hemisphere. 
Because the
sphere is compact, a large $d$ jump from what appears to be an 
outer edge of a lawn, in any direction, could return the grasshopper
to the lawn. For small $d$, however, the construction of Lemma $2$ 
suggests that the hemisphere is not optimal, and it seems plausible
that cogwheel-type solutions will again emerge.
This does not, however, give any definitive insight into the solutions
of versions of the problem relevant to Bell inequalities, since these
require (at least) the antipodal condition as an additional
constraint.   The construction of Lemma $2$ does not generalise to 
a construction of an antipodal lawn on the sphere, at least in any
straightforward way, and our numerical 
cogwheel solutions do not generally appear to either.
It thus remains plausible that the antipodal condition is significantly
relevant and that the hemisphere is optimal for small $d$
among antipodal lawns. Now that our numerical techniques are
well tested in the planar case, it should be simple to transfer
them to the spherical case and thus resolve this question.

There are many other questions related to and generalizations of the grasshopper problem that 
it would be interesting to address. We list some of these here. 

\qquad{\bf Restrictions on the lawn:} Interesting
restrictions include requiring  
that the shapes be connected, or simply connected, or convex. 
Our results suggest that each of the last three is a genuine
restriction.\footnote{Note that, if the optimal shape is disconnected,
there will exist sequences of connected shapes with success 
probability tending to the optimal probability.   
Such shapes can be constructed from the optimal shape by
subtracting small areas from the optimal lawn and then 
connecting all its disconnected parts by suitably thin
strips.}   
The optimal shape appears to be non-convex 
for any value of $d$, and not connected for large $d$. 

\qquad{\bf Variable density lawns:} It seems
natural to allow lawns of less than unit density, but unclear 
whether such lawns might actually be optimal for any value
of $d$.  That is, does there exist a jump distance $d$ with an optimal
probability density $\mu$ such that the set 
$\{ \mathbf{r} \, : \, \mu( \mathbf{r} ) = 1 \}$ has
measure less than one? 
Our results to date are consistent with the optimal 
density always being an identity function: that is, 
with the density of the optimal lawn always being one or zero.  

The spin models, which give numerical
evidence about the form of optimal lawn configurations, also effectively assume unit
density lawns at their finest scale of resolution. There are alternatives to this discretization scheme. One could, for
instance, allow intermediate spin values between 0 and 1. 
This may not be necessary, since it seems plausible that a simulation of the regular
discretized model could effectively 
produce such a density function without allowing intermediate spin
values. 
For example, a checkerboard lawn region is a simple discretized 
model of a lawn region with $\mu = \frac{1}{2}$. 
Alternatively, if the small-scale checkerboard regularities turn out
to have a significant effect, a continuum lawn region with $\mu =
\frac{1}{2}$
could be modelled by
a discretized region in which the grid cells are independently randomly
and equiprobably filled ($s_i =1$) or empty ($s_i =0$). 
More generally, a continuum lawn region with density $\mu$ 
could be modelled by a discretized lawn region in which a proportion $\mu$ of
the cells are randomly filled. One might thus
hope that the existing discretized model would identify optimal
lawns with non-uniform density, if any exist. Our numerical 
results so far do not suggest any optimal lawns of this type. 
Finer scale discretizations would allow further investigation 
of this question. 

\qquad{\bf Generalisations to other dimensions:} The problem generalises
to ${\mathbb R}^n $ for any positive integer $n$. The case $n=1$ is easy to solve. Consider the lawn 
defined by the segments $\lbrack 0, \frac{1}{N} \rbrack$,
$\lbrack d , d+ \frac{1}{N} \rbrack$, $\ldots$, 
$\lbrack (N-1) d , (N-1) d+ \frac{1}{N} \rbrack$.
This has total length $1$ and gives a probability 
$\frac{N-1}{N}$ of the grasshopper remaining on the lawn. 
The sequence of such lawns for positive integer $N$ thus gives us the supremum 
value $1$ for the probability. This class of solutions, whose members are periodic over increasingly large ranges
and have success probabilities increasingly close to $1$,
appears to have no
analogy in higher dimensions, where the problem seems much harder.  
Intuitively, this appears to be because of a mismatch between the dimensionality
of the jump and the space in which it takes place.    

\qquad{\bf Generalisations to other spaces:}  The grasshopper problem 
also seems interesting and natural on other Riemannian manifolds. 
One example is the surface of the sphere $S^n$ -- and indeed 
the problem was originally motivated by the case of $S^2$. 
The grasshopper problem seems particularly natural on the sphere,
as on the plane, since jumps of fixed distance in any direction
are related by the sphere's rotational symmetry.
Another interesting example is the torus $T^n$, obtained by identifying the edges
of a hypercube in the usual way.   While this has a lower degree of symmetry
than the sphere, it lends itself naturally to discretization by
lattices with periodic boundary conditions.  For both the sphere
and torus the solution must in general depend on the ratio of the lawn volume 
to the manifold volume. So, if we keep the convention that the lawn
has volume one then we should consider the manifold volume 
(or equivalently the sphere radius or the hypercube edge length) as a 
second variable in the problem along with $d$.   

\qquad{\bf Generalisations to other metrics:} Rather than using
  Euclidean distance (the $L_2$ norm) we can consider a different metric,
  for instance the Manhattan metric defined by the $L_1$
    norm.  The latter is particularly
  interesting for the discrete version of the grasshopper problem on a
  square grid, which can be studied with a spin model. In this case
  there is no error in the distance measurement associated with the
  discretization of the system. Alternative grids, for instance
  hexagonal, can also be studied with the appropriate definitions of
  Manhattan distance.

\qquad{\bf Extensions of the problem:}  The grasshopper problem can be naturally extended
in a number of ways. One variation, which is motivated by our original Bell inequality
problem \cite{kent2014bloch}, is to allow two independently defined lawns,
using two different species of lawn seed that can coexist in the same region, with each independently allowed any density between zero and one inclusive. In this version of the problem the grasshopper lands
randomly on the first lawn and jumps as above. The aim is to find
a pair of lawn configurations that maximises the probability that it ends up on
the second lawn.

Other variations of the single lawn grasshopper problem allow $N$ independent random jumps of distance
$d$, either simply requiring that the grasshopper is on the lawn
after the $N$-th jump, or else imposing the condition that it must
also be on the lawn after each jump. 
One could also allow a variable random jump distance, for example
Poisson distributed with mean $d$.  

Another interesting variant is the {\it ant problem}, which
is defined for lawns of unit density.  Like
the grasshopper, the flying ant arrives at a random point on the lawn.
It then sets out to walk a distance $d$
in a random direction. Alas, however, if at any point it leaves the lawn, 
it dies. It seems clear that disconnected 
lawn shapes will not be optimal for any $d$ in this case. Moreover, it is not
evident that the shapes optimal for the grasshopper problem are
optimal for the ant problem for any $d$.

A further alternative is to model the grasshopper (or the ant) when
external forces influence its trajectory.  For example, we can
define a simple model of the grasshopper in a breeze by supposing
that a fixed vector is added to the  
random length $d$ vector defining its regular jump.\footnote{The breeze could be
modelled more realistically by modelling an actual jump trajectory 
with fixed initial angle of elevation and speed, taking into account
the breeze force and gravity. Another interesting variation 
along these lines is to model the grasshopper's trajectory
under gravity (not necessarily with any breeze) when
jumping on an inclined plane or other embedded manifold.}
It would also be interesting to understand the 
properties of the corresponding spin models, which 
describe an anisotropic interaction whose range 
is fixed in any given direction but varies with the direction.

\qquad{\bf Applications to catalytic reactions:} One source of
motivation for several of the variations above
is that the grasshopper
problem defines a simple model of catalytic reactions
that give significant impetus to the catalyst.  
For example, consider a chain reaction in which
a randomly selected nucleus of an atom of element $A$ is
impacted by a high energy particle $P$ and
undergoes fission. Suppose that the fission process produces
a further particle $P$ of fixed energy, independent of the energy 
of the original particle $P$, which travels a 
distance $d$ before becoming near stationary. 
At this point, if surrounded by atoms of the same type $A$, 
it will be absorbed by another nucleus, again causing fission, and so
on. Suppose that there
are a finite number $N$ of atoms of type $A$ contained in a region $R$. 
Suppose also that the region $R$ is surrounded by atoms of element $B$
that slow the particles at the same rate as those of element $A$, so that the particles 
travel the same distance $d$ whether they remain in $R$ or not.
However, if they come to rest surrounded by type $B$ atoms, they 
undergo no further reactions. A reaction of this type can be modelled 
by the grasshopper problem in the relevant number of 
dimensions (three if the region $R$ is unconstrained,  
two if it is constrained to be a thin solid, with fixed 
height $h \ll d$ and constant cross-section).      
The maximum initial reaction rate arises when the 
region $R$ solves the grasshopper problem. 

\qquad{\bf Applications to morphogenesis:}  Our numerical 
solutions strikingly illustrate the emergence of structures with
discrete symmetries from an isotropic problem with continuous 
rotational symmetry. They also bear at least a 
passing resemblance to patterns seen elsewhere in nature, 
including the contours of flowers, the patterns seen 
on some seashells and the stripes on some animals. 

Turing's well-known theory of morphogenesis \cite{turing1952chemical} hypothesises that
many such natural patterns arise as solutions to reaction-diffusion
equations.   This possibility has been 
demonstrated experimentally \cite{tompkins2014testing}. 
Our results suggest that a rich variety of pattern formation can
also arise in systems with effectively fixed-range interactions, 
including interactions associated with the sort of catalytic reaction
described 
above. It may be worth looking for explanations of this type in any context
where highly regular patterns naturally arise and are not 
otherwise easily explained. 

\qquad{\bf Further applications to quantum information theory:}
Our investigation was motivated by analysing specific Bell
inequalities that distinguish the predictions of quantum 
theory for a two qubit singlet state from those of local hidden
variable theories. These particular inequalities have an
especially simple geometric interpretation which generalises
to the plane and other surfaces.   

Spin models amenable to simulated annealing and parallel tempering 
could also be used to investigate other types of Bell inequality
for pairs of qubits and higher dimensional entangled states.
It would be very interesting, for example, to apply them
to the problem of finding the full ranges of Werner 
states \cite{Hirsch2017betterlocalhidden} 
that can be simulated by local hidden variable models,
and to explore similar questions for other interesting entangled
states, including multipartite states.
  
Nor are our techniques restricted to analysing Bell inequalities.   
A local hidden variable model is effectively a classical
algorithm for simulating quantum theory, and our motivating
goal is to find the best such classical algorithm according
to a natural metric.
Simulated annealing and parallel
tempering methods can also be applied to search for more
general classical algorithms that optimize a quantity of interest.
In particular, they could be applied to other problems 
of interest in quantum information theory, including 
problems of quantum communication complexity. 
One relatively simple example is the problem of finding algorithms
that simulate quantum
correlations with shared randomness and finite one-way classical
communication \cite{PhysRevLett.91.187904}.   
For projective measurements on pairs of qubits, such
algorithms can be thought of as generalised grasshopper
models, in which Alice has a single lawn, Bob has a set of available 
lawns, and Alice chooses which of Bob's lawns is used in each given run.

\section*{Acknowledgements}
We are grateful to Chris Amey, Simon Lentner, Jonathan Machta and Mykhaylo Tyomkyn for interesting and helpful discussions.
This work was supported by an FQXi grant and by Perimeter Institute
for Theoretical Physics. Research at Perimeter Institute is supported
by the Government of Canada through Industry Canada and by the
Province of Ontario through the Ministry of Research and Innovation. OG is funded by the NSF under the Grant No. PHY-1314735.

\section*{References}

\bibliographystyle{unsrtnat}
\bibliography{grasshopper}
\end{document}